\documentclass[superscriptaddress,twoside,twocolumn,showpacs,preprintnumbers,amssymb,amsmath,pre]{revtex4}
\usepackage{graphicx}
\usepackage{dcolumn}
\usepackage{bm}
\begin{document}
\title{Implications of Realistic Fracture Criteria on Crack Morphology} 
\author{Bj{\o}rn Skjetne}
 \affiliation{Department of Physics,
              Norwegian University of Science and Technology,\\
              N-7491 Trondheim, Norway}
\author{Alex Hansen}
 \affiliation{Department of Physics,
              Norwegian University of Science and Technology,\\
              N-7491 Trondheim, Norway}
\date{\today}
\begin{abstract}
We study the effects realistic fracture criteria have on 
crack morphology obtained in numerical simulations with a stochastic
discrete element method. 
Results are obtained with two criteria which are consistent with
the theory of elasticity and compared
with previous results using the
original criterion, chosen when the method was first published.
The conventional choice has been to consider the combined loading
as an interaction between bending and tensile forces only, leaving
out shear forces altogether. Moreover the combination of bending
and tension used in the old criterion is correct only for plastic
deformations. Our results show that
the inclusion of shear forces have a profound effect on crack
morphology. We consider two types of external loading, torsion
applied to a circular cylinder and tension applied to a cube.
In the tensile case, the exponent which
characterises scaling of crack roughness with system size
is found to be very close to the experimental value $\zeta\sim0.5$
when realistic fracture criteria are used. In the present
calculations we obtain $\zeta=0.52$, a value which remains constant
for all disorders. It is proposed that the small-scale exponent 
$\zeta=0.8$ appears as a consequence of cleavage between crystal
planes and consequently requires a different fracture criterion than
that which is used on larger scales.
\end{abstract}
\maketitle

\section{Introduction}
\label{intro}
Material properties have long been studied using finite
element methods.
A different approach to fracture and breakdown
phenomena was introduced almost three decades ago within the
statistical physics community.
In this approach, a macroscopic material is though to be made
up of discrete elements arranged on a lattice or grid.
Into this discretized version of the material
random variations in structural properties are
introduced at the scale of the discrete elements. This can be
done for the elastic properties of the elements or, as is
more common, it can be made to affect the individual breaking
strengths of the elements.
The resulting breakdown process, whether it takes the form
of electrical failure in a network of fuses or describes
the elastic breaking of a
discretized continuum, is complex and results in a rough
crack interface.

In stochastic discrete element modeling of elastic media there
have been mainly two ways to model forces within
the continuum, i.e., in terms of 'springs'~\cite{born} or in 
terms of 'beams'~\cite{roux,herr}. 
The former approach is simpler and less requiring of
computational resources since in this case bending and shear
forces are absent on the level of the individual element
(the spring). The latter approach is more realistic since it
reproduces the full mechanical response of a real solid, i.e.,
each 'beam' transmits axial forces, bending moments,
transverse shear and torsion to its adjoining neighbours.

One popular quantity to study has been crack morphology as 
quantified by the roughness exponent. This can be measured 
experimentally and is therefore an important quantity to be 
reproduced by the theoretical models available. The typically rough 
crack surface that is obtained in materials with a heterogeneous 
microstructure is due to a complex interplay between stresses and 
local variations in material structure. Such variations can be
due to a granular structure with different grain sizes, with the 
grains being randomly distributed and subject to different bonding 
strengths. Material disorder can be due to a fibrous structure, or 
a structure characterized by pores and voids, or it can be 
dominated by the presence of microscopic cracks, inclusions and 
fault lines. Strength variations may be in the form of weak spots 
in the material or local regions that are stronger than the 
surroundings. 

If we consider fracture in such disordered media, this is a coupled
process whereby stresses evolve according to how cracks grow while 
cracks develop according to how stresses are distributed. At some 
point the fracture process goes from being disorder dominated, where 
new cracks appear randomly, to being stress dominated, where smaller 
cracks merge into a large crack. In this coupled process a path is 
now forged through the medium by the moving front of this crack. The
fracture criterion plays an important role in determining the exact 
nature of the path taken. In other words, its role is to decide the 
outcome of the interplay between stress and disorder. 
It is therefore extremely important to study how different
fracture criteria influence the fracturing process. 
As we shall see the role played by the fracture criterion
is also affected by, and 
intimately associated with, lattice morphology. A criterion which 
does not allow for failure in transverse loading will display a 
tendency towards fracturing along lattice planes.

In simple stochastic models of fracture, such as the random fuse
model, there isn't much choice as far as the fracture criterion
is concerned. Here the ratio of the current which flows through
an element to the burn-out threshold of that element is what
constitutes the breaking criterion. There really is no other
choice. In elastic fracture, on
the other hand, there are several modes of loading and each of
these can contribute to the breaking of an element. This is
especially so whenever forces are defined in terms of 'beams'.
In the case of 'springs' only axial forces exist on the scale
of the individual element. When the full elastic response is 
included, however, an element breaks if the axial load exceeds a 
certain limit or if the shear exceeds a certain limit.
In general the situation is a combined loading.

Much effort has been expended
within engineering communities to obtain simplified, yet
realistic, interaction formulae relevant to various loading
conditions. Typically, the nature of these fracture criteria
depend on the type of material used, the shape or cross section
of the element involved, and the specific application concerned. 
They can be theoretically derived or empirically deduced from 
laboratory tests, or they can be based on a combination of 
approaches. 

Stochastic fracture models were developed 
within the statistical physics community and consequently much 
interest has been focused on the complex process of interaction 
between stress and disorder. This is probably one reason why
failure criteria have received less attention than what would have 
been the case within the engineering community.
\begin{figure}
\centering
\includegraphics[scale=0.27]{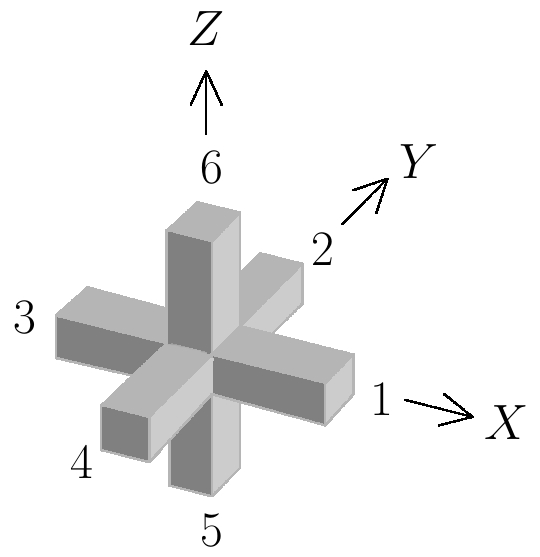}
\caption{Enumeration scheme for the discrete 'beam' elements of 
         a cube lattice connecting node~$i$ to its nearest 
         neighbours~$j=1$ to~$j=6$, showing the coordinate system 
         with node $i$ as its origin.
         \label{6beam}}
\end{figure}

\section{Discrete Element Model}
\label{secmodel}
Our model is a deformable lattice in the form of
a regular cube with size $L\times L\times L$, where
each node is connected to its nearest neighbours
by linearly elastic beams.
Forces acting on the nodes have been deduced from the effect 
a concentrated end-load has on a beam with no 
end-restraints~\cite{roar,cran}. A coordinate system is placed 
on each node, and the enumeration of the connecting beams 
follows an anti-clockwise 
scheme within the $XY$-plane, i.e., beginning with the beam which 
lies along the positive $X$-axis and ending with that
which extends upwards along the positive $Z$-axis, 
see Fig.~\ref{6beam}.

At each stage of the breaking process, the updated displacements
for each node is obtained from
\begin{eqnarray}
    \sum_{j}{\bf D}_{ij}
             \left[\begin{array}{l}
                      x_{i}\\
                      y_{i}\\
                      z_{i}\\
                      u_{i}\\
                      v_{i}\\
                      w_{i}\\
                    \end{array}
             \right]=\lambda
             \left[\begin{array}{l}
                      X_{i}\\
                      Y_{i}\\
                      Z_{i}\\
                      U_{i}\\
                      V_{i}\\
                      W_{i}\\
                   \end{array}
             \right],
              \label{ma3x}
\end{eqnarray}
which is solved iteratively via relaxation using the conjugate 
gradient method~\cite{hest,teuk}. 
This minimizes the 
elastic energy to obtain those displacements for which the
sum of forces and moments on each node vanish, i.e. the
mechanical equilibrium.
In Eq.~(\ref{ma3x}), $x_{i}$, $y_{i}$ and $z_{i}$ are the
coordinate displacements of node~$i$ relative to its 
starting position before fracturing is initiated. Likewise,~$u_{i}$,
$v_{i}$ and $w_{i}$ are the angular displacements around the
$X$-, $Y$- and $Z$-axes, respectively (see Fig.~\ref{6beam}).

Presently we use the same expressions for force and moment 
as those used in Refs.~\cite{herr} and~\cite{skje}. We thus
have three constants 
\begin{eqnarray}
    \alpha=\frac{\ell}{EA},\qquad
    \beta=\frac{\ell}{GA},\qquad
    \gamma=\frac{\ell^{3}}{EI},
    \label{matcons}
\end{eqnarray}
where $E$ and $G$ are Young's modulus and the shear modulus, 
respectively, $A$ is the area of the discrete element cross section,
$\ell$ is its length and $I$ the moment of inertia
about the centroidal axis. Our choice of these
parameters mirrors that of Refs.~\cite{herr} and~\cite{skje}, i.e.,
the length is set to~$\ell=1$ and we use $\alpha=1$,
$\beta=30/7$ and $\gamma=60/7$. 
Additionally, we define the quantity
\begin{eqnarray}
    \rho=\frac{\ell}{JG},
    \label{epscons}
\end{eqnarray}
where $J$ is the moment of inertia for torsion. Here we use $\rho=1$.
In the following we define 
\begin{eqnarray}
	\delta x_{j}\equiv x_{i}-x_{j}, 
\end{eqnarray}
and similarly for the other five coordinates.

Six terms contribute to each of the force components in Eq.~(\ref{ma3x}). 
For instance, if we imagined the neighbouring nodes to be fixed, a
translation $x_{i}$ of the central node~$i$ would induce axial 
forces in beams~$1$ and~$3$ and transverse forces in 
beams~$2$,~$4$,~$5$ and~$6$. 
If we take into account the displacements of the neighbouring
nodes as well, the axial force on node~$i$ from beam~$1$ is
\begin{eqnarray}
	A_{i}^{(1)}=-\frac{1}{\alpha}\delta x_{1},
          \label{xFor}
\end{eqnarray}
while the transverse force on node~$i$ from beam~$2$, along the
$X$-axis, is given by
\begin{eqnarray}
          {_{x}}S_{i}^{(2)}=-\frac{1}{\beta+\frac{\gamma}{12}}
          \Bigl[\delta x_{2}
         -\frac{1}{2}\bigl(w_{i}+w_{j}\bigr)\Bigr].
          \label{yFor}
\end{eqnarray}
In each case~$j$ refers to the neighbour depicted 
in Fig.~\ref{6beam}. Consequently, 
\begin{eqnarray}
    X_{i}=A_{i}^{(1)}+A_{i}^{(3)}
         +\sum_{j\ne1,3}^{6}{_{x}}S_{i}^{(j)}
           \label{xcomp}
\end{eqnarray}
is how the force on node~$i$ along the $X$-axis
depends on the displacements and rotations of its 
six nearest neighbouring nodes.
Similar expressions are deduced for $Y_{i}$ or $Z_{i}$ by considering
translations along the $Y$-axis or the $Z$-axis, respectively.

An angular displacement $u_{i}$ about the $X$-axis 
with the neighbouring nodes fixed would create torque in
beams~$1$ and $3$, and bending in beams $2$, $4$, $5$ and $6$. 
More generally, the torque in node~$i$ from beam~$1$ is
\begin{eqnarray}
    T_{i}^{(1)}=-\frac{1}{\rho}\delta u_{1},
          \label{xTor}
\end{eqnarray}
while the bending moment from beam~$2$ is
\begin{eqnarray}
	{_{u}}M_{i}^{(2)}\hspace{-0.5mm}
                    =-\frac{1}{\beta+\frac{\gamma}{12}}
                       \Bigl[\frac{\beta}{\gamma}\delta u_{2}
			       \hspace{-0.5mm}+\hspace{-0.5mm}
			       \frac{1}{2}\bigl(\delta z_{2}
			       \hspace{-0.5mm}+\hspace{-0.5mm}
                               \frac{2}{3}u_{i}
			       \hspace{-0.5mm}+\hspace{-0.5mm}
			       \frac{1}{3}u_{j}
                      \bigr)\Bigr].
          \label{mFor}
\end{eqnarray}
For the angular force on node~$i$ about the $X$-axis
and its dependence on the displacements 
of the six neighbouring nodes, we have
\begin{eqnarray}
    U_{i}=T_{i}^{(1)}+T_{i}^{(3)}
         +\sum_{j\ne1,3}^{6}{_{u}}M_{i}^{(j)},
           \label{ucomp}
\end{eqnarray}
now with similar expressions for $V_{i}$ and $W_{i}$.

To express the thirty-six force components in Eq.~(\ref{ma3x}) more
compactly,
\begin{eqnarray}
    r_{j}=\prod_{n=0}^{j-1}\bigl(-1\bigr)^{n}
\end{eqnarray}
and
\begin{eqnarray}
    s_{j}=\bigl(-1\bigr)^{j}r_{j}
\end{eqnarray}
are quantities which we define for notational convenience, to keep 
track of the signs and contributions from neighbouring beams. 
The~$j$ in each case refers to the neighbouring beams as
shown in Fig.~\ref{6beam}. The 
Kronecker delta, moreover, has been used to construct
\begin{eqnarray}
    \widehat{\lambda}_{s,t}=
     \delta_{sj}+\delta_{tj},
      \label{inc}
\end{eqnarray}
i.e., an operator which includes $s$ and $t$ in the sum over 
neighbours (excluding the other four), and
\begin{eqnarray}
    \widehat{\chi}_{s,t}=
     \bigl(1-\delta_{sj}\bigr)\bigl(1-\delta_{tj}\bigr),
      \label{exc}
\end{eqnarray}
which instead excludes $s$ and $t$ from the sum over neighbours
(including the other four). 

For the six components making up the force on node~$i$ along
the $X$-axis, i.e., Eq.~(\ref{xcomp}), we can now state this
on a compact form as
\begin{widetext}
\begin{eqnarray}
    X_{i}=-\frac{1}{\alpha}
            \sum_{j=1}^{6}
	     \widehat{\lambda}_{1,3}\delta x_{j}
          -\frac{1}{\beta+\frac{\gamma}{12}}
            \sum_{j=1}^{6}
	     \widehat{\chi}_{1,3}
	      \biggl\{
	       \delta x_{j}
              +\frac{r_{j}}{2}
         \Bigl[\widehat{\lambda}_{5,6}\bigl(v_{i}+v_{j}\bigr)+
               \widehat{\lambda}_{2,4}\bigl(w_{i}+w_{j}\bigr)
         \Bigr]
             \biggr\},
	      \label{forceXi}
\end{eqnarray}
and $Y_{i}$ as
\begin{eqnarray}
    Y_{i}=-\frac{1}{\alpha}
            \sum_{j=1}^{6}
  	     \widehat{\lambda}_{2,4}\delta y_{j}
          -\frac{1}{\beta+\frac{\gamma}{12}}
            \sum_{j=1}^{6}
	     \widehat{\chi}_{2,4}
              \biggl\{
     	       \delta y_{j}
              +\frac{r_{j}}{2}
         \Bigl[\widehat{\lambda}_{5,6}\bigl(u_{i}+u_{j}\bigr)+
               \widehat{\lambda}_{1,3}\bigl(w_{i}+w_{j}\bigr)
         \Bigr]
             \biggr\}.
	      \label{forceYi}
\end{eqnarray}
In the same way, $Z_{i}$ becomes
\begin{eqnarray}
    Z_{i}=-\frac{1}{\alpha}
            \sum_{j=1}^{6}
	     \widehat{\lambda}_{5,6}\delta z_{j}
          -\frac{1}{\beta+\frac{\gamma}{12}}
            \sum_{j=1}^{6}
	     \widehat{\chi}_{5,6}
	      \biggl\{
	       \delta z_{j}
              -\frac{s_{j}}{2}
         \Bigl[
               \widehat{\lambda}_{2,4}\bigl(u_{i}+u_{j}\bigr)+
               \widehat{\lambda}_{1,3}\bigl(v_{i}+v_{j}\bigr)
         \Bigr]
             \biggr\},
	      \label{forceZi}
\end{eqnarray}
Next, Eq.~(\ref{ucomp}) for angular displacements about the
$X$-axis is written out in full as
\begin{eqnarray}
    U_{i}=-\frac{1}{\rho}
            \sum_{j=1}^{6}
	     \widehat{\lambda}_{1,3}\delta u_{j}
          -\frac{1}{\beta+\frac{\gamma}{12}}
            \sum_{j=1}^{6}
             \widehat{\chi}_{1,3}
           \biggl\{
	       \frac{\beta}{\gamma}\delta u_{j}
               +\frac{r_{j}}{2}
           \Bigl[\hspace{0.5mm}
		  \widehat{\lambda}_{5,6}\delta y_{j}
	         -\widehat{\lambda}_{2,4}\delta z_{j}
           \Bigr]
          +\frac{1}{3}\bigl(u_{i}+\frac{1}{2}u_{j}\bigr)
           \biggr\},
	    \label{forceUi}
\end{eqnarray}
and $V_{i}$, for angular displacements about the
$Y$-axis, becomes
\begin{eqnarray}
    V_{i}=-\frac{1}{\rho}
            \sum_{j=1}^{6}
	     \widehat{\lambda}_{2,4}\delta v_{j}
          -\frac{1}{\beta+\frac{\gamma}{12}}
            \sum_{j=1}^{6}
             \widehat{\chi}_{2,4}
           \biggl\{
	       \frac{\beta}{\gamma}\delta v_{j}
               +\frac{r_{j}}{2}
           \Bigl[\hspace{0.5mm}
		  \widehat{\lambda}_{5,6}\delta x_{j}
		 -\widehat{\lambda}_{1,3}\delta z_{j}
           \Bigr]
          +\frac{1}{3}\bigl(v_{i}+\frac{1}{2}v_{j}\bigr)
           \biggr\}.
	    \label{forceVi}
\end{eqnarray}
Lastly, for angular displacements about the $Z$-axis, we get
\begin{eqnarray}
    W_{i}=-\frac{1}{\rho}
            \sum_{j=1}^{6}
	     \widehat{\lambda}_{5,6}\delta w_{j}
          -\frac{1}{\beta+\frac{\gamma}{12}}
            \sum_{j=1}^{6}
             \widehat{\chi}_{5,6}
           \biggl\{
	       \frac{\beta}{\gamma}\delta w_{j}
               +\frac{r_{j}}{2}
           \Bigl[\hspace{0.5mm}
	          \widehat{\lambda}_{2,4}\delta x_{j}
		 +\widehat{\lambda}_{1,3}\delta y_{j}
           \Bigr]
          +\frac{1}{3}\bigl(w_{i}+\frac{1}{2}w_{j}\bigr)
           \biggr\}.
	    \label{forceWi}
\end{eqnarray}
\end{widetext}

\section{Disorder}
\label{disrd}
To include structural disorder we
generate a random number~$r$ on the unit 
interval $[0,1]$ and let this represent the cumulative 
threshold distribution. We assign thresholds according to 
$t_{\rm F}=r^{D}$, where $D>0$ is a power law with a maximum
threshold of~$1$ and a tail extending towards zero. 
The cumulative distribution function is then given by
\begin{equation}
    P(t_{\rm F})=t_{\rm F}^{1/D},
       \label{posd}
\end{equation} 
where $0\le t_{\rm F}\le 1$. 
The case of $D=0$ corresponds to all thresholds being the 
same ($t_{\rm F}=1$), i.e., we have a homogeneous medium without 
structural disorder. An increase in the magnitude of the 
exponent~$|D|$ causes the coefficient of variation with respect 
to any two random numbers $r$ and $r^{\prime}$
on the interval~$[0,1]$ to increase. 
Therefore large values of $|D|$ correspond to strong disorders and
small values to weak disorders. 

\section{Fracture Criteria}
\label{fcrit}
The original fracture criterion introduced by 
Herrmann et al. in Ref.~\cite{herr}
considers a combination of bending and axial force,
where beams fail when
\begin{equation}
    \left(\frac{F}{t_{F}}\right)^{\hspace{-1mm}2}+
     \frac{|M|}{t_{M}}>1,
      \label{f2m}
\end{equation}
that is, using a squared term for the axial force and a linear term 
for the bending moment. The quantities $t_{\rm F}$ and $t_{\rm M}$
are thresholds for the amount of bending the element can support
before failing.
In applications other than stochastic modeling,
there are two scenarios where this particular
fracture criterion is frequently used. One is in connection with combined
loadings for slender beams in compression~\cite{stal}. The other
is for materials where plastic yielding occurs, in which case the
loading can be either tensile or compressive~\cite{chak}. Presently
we consider brittle fracture only.

\subsection*{Combined Axial Force and Bending}
In wood constructions the region of safe loading for beam columns 
with rectangular cross sections, when subjected to a combination of 
axial tension and bending~\cite{ndsw}, is given as
\begin{equation}
    \frac{F}{t_{F}}+
     \frac{M}{t_{M}}>1
      \label{wood_t_u}
\end{equation}
in the unixial case, and
\begin{equation}
    \frac{F}{t_{F}}+
    \frac{M_{x}}{t_{M_{x}}}+
    \frac{M_{y}}{t_{M_{y}}}>1
       \label{wood_t_b}
\end{equation}
in the biaxial case, while the criterion for 
failure in compression is
\begin{equation}
	\left(\frac{F}{t_{F}}\right)^{\hspace{-1mm}2}\hspace{-0.5mm}+
     \frac{M}{t_{M}}>1
      \label{wood_c_unax}
\end{equation}
in the uniaxial case, and
\begin{equation}
	\left(\frac{F}{t_{F}}\right)^{\hspace{-1mm}2}\hspace{-0.5mm}+
     \frac{M_{x}}{t_{M_{x}}}+
      \frac{M_{y}}{t_{M_{y}}}>1
      \label{wood_c_biax}
\end{equation}
in the biaxial case. Deflection of the beam in the
presence of a compressive force tends to magnify the moment that
causes it, and consequently more emphasis is lent to the axial term.
This distinction between tensile and compressive
loading, however, is irrelevant to applications in the discrete 
element model. The reason is that the model 
is meant to describe a {\em continuum} rather than a physical 
lattice. In order to emulate the behaviour of a continuum, 
elements defining forces between nodes should not be considered
to be slender. In fact, they should not in any way buckle 
within the structure of the material! Interaction formulas
relevant to compression should therefore have the same functional 
form as those relevant to tension.

\begin{figure} [t]
\centering
\includegraphics[angle=0,scale=0.18]{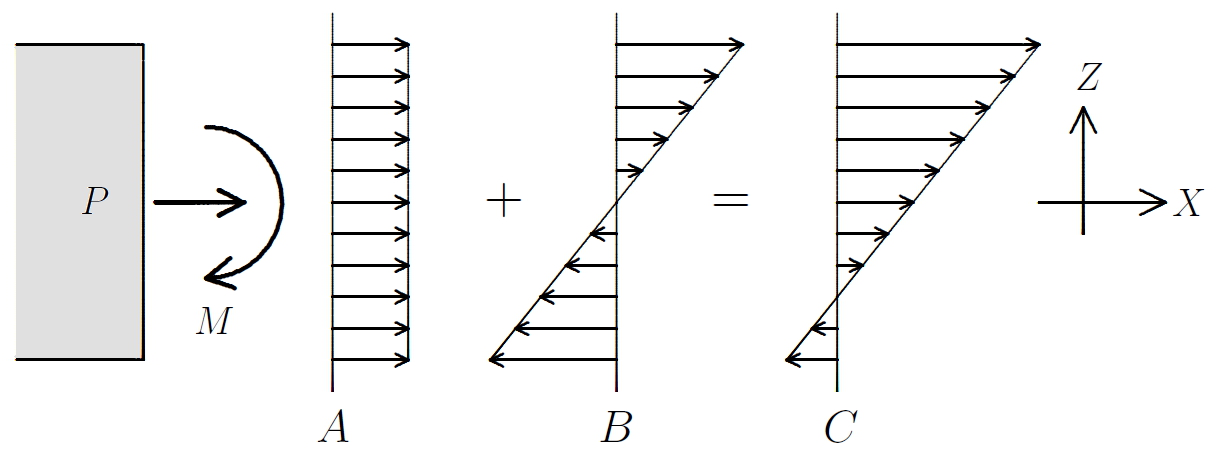}
\caption{Superposition of stress in a
	 beam subject to combined axial load~$P$ and bending 
	 moment~$M$. The cross-sectional area is~$A$, on which 
	 the stress distribution 
	 due to $P$ alone is shown in~(A), that
	 due to $M$ is shown in~(B),
	 and the superposition of the two is shown in~(C).
         \label{tensbend}}
\end{figure}
This choice is easy to justify using standard elastic theory.
In the two-dimensional case, we use the superposition principle 
for combined loadings. For a beam with its axis lying along the 
$X$-axis, and with a cross-section perpendicular to this, the normal
stress caused by axial loading in the direction of the positive 
$X$-axis (tension) is given by
\begin{equation}
    \sigma_{x,t}=\frac{P}{A},
    \label{sigxax}
\end{equation}
where $P$ is the force and~$A$ is the cross-sectional area, 
see~Fig.~\ref{tensbend}A. Normal stresses also arise in bending.
Assuming the beam is bent within the $XZ$-plane (upper surface tensile)
these stresses are
\begin{equation}
    \sigma_{x,b}=-\frac{Mz}{I}
    \label{mzi}
\end{equation}
where the bending moment
is~$M$ and $I$ is the moment of inertia of the cross-sectional
area about the neutral axis. Normal stresses 
are seen to depend linearly upon the vertical distance~$z$ from the
neutral axis, see Fig.~\ref{tensbend}B.
Adding Eqs.~(\ref{sigxax}) and~(\ref{mzi}) we get
\begin{equation}
    \sigma_{x}=\sigma_{x,t}+\sigma_{x,b}
    \label{sigtot}
\end{equation}
for the normal stress of the combined loading, that is,
\begin{equation}
    \sigma_{x}=\frac{P}{A}-\frac{Mz}{I},
    \label{totsig}
\end{equation}
and the maximum~$|\sigma_{x}|$ occurs along the top surface of
the beam, as can be seen from~Fig.~\ref{tensbend}C. 
In contrast, below the neutral axis (negative~$z$)
the normal stress due
to the axial load is reduced since Eq.~(\ref{mzi}) becomes
negative here. It is at its lowest along the bottom 
surface.
If we reverse the sign on~$P$ and consider compressive axial
forces, we find~$|\sigma_{x}|$ to be largest along the bottom
surface for a beam bent like this. 

A positive moment is defined as one where the beam is concave up,
i.e., with the bottom surface in tension.
Hence, with $z=-c$
representing the outermost fibre on the cross section,
\begin{equation}
    \sigma=\frac{F}{A}+\frac{Mc}{I},
    \label{FAMcI}
\end{equation}
is the combined stress of axial tension and bending at this 
particular location. Dividing through Eq.~(\ref{FAMcI}) by the 
maximum value of the normal stress within the elastic 
range,~$\sigma_{\rm p}$, we obtain
\begin{equation}
    \frac{\sigma}{\sigma_{\rm p}}=\frac{F}{F_{\rm y}}
                                 +\frac{M}{M_{\rm y}},
    \label{divThru}
\end{equation}
where 
\begin{equation}
    F_{\rm y}=\sigma_{\rm p}A
    \label{fypa}
\end{equation}
is the axial force at its elastic limit, and
\begin{equation}
	M_{\rm y}=\frac{I}{c}\hspace{0.5mm}\sigma_{\rm p}
    \label{mypa}
\end{equation}
is the bending moment at its elastic limit.
Modeling a material which cannot deform plastically, i.e., which
fails beyond the elastic limit, we can then identify~$F_{\rm y}$
and~$M_{\rm y}$ as breaking thresholds in~$F$ and~$M$. 
If these thresholds are denoted~$t_{F}$ and~$t_{M}$,
one has to remove from the
calculations those elements for which 
\begin{equation}
    \sigma>\sigma_{\rm p},
    \label{sigLsig}
\end{equation}
and therefore, according to Eq.~(\ref{divThru}), we must remove those 
elements for which the combination of~$F$ and~$M$ are such that
\begin{equation}
    \frac{F}{t_{F}}+\frac{M}{t_{M}}>1
    \label{fcsigsig}
\end{equation}

In our calculations we will not be interested in the details of
{\it where} a 'beam' is most stressed. What we require is to identify the
maximum stress that occurs in a combined loading. In the case of
axial tension Eq.~(\ref{wood_t_u}) selects those beams where the most
stressed material fibre is beyond the 
breaking threshold (this will be on the convex side of the beam,
be that on the upper or lower surface).
In the case of compression, a beam in the same bent configuration
would be most stressed on the opposite surface (the concave side),
and this maximum stress is still given by Eq.~(\ref{wood_t_u}).
We therefore need not distinguish between the convex and the concave sides
in our application. In the discrete element model each element is
either kept or removed depending on the magnitude of its greatest
combined stress. The sign on~$F$ or~$M$ then becomes 
irrelevant.

For our purposes, then, elements that fail can be identified in 
three dimensions using
\begin{equation}
    \frac{|F|}{t_{F}}+\frac{|M|}{t_{M}}>1,
    \label{FMM2}
\end{equation}
where the biaxial moment is given by
\begin{equation}
    M=\sqrt{M_{x}^{2}+M_{y}^{2}}
    \label{MMMxy}
\end{equation}
and the same thresholds $t_{\rm M}$ apply in all planes of bending.

\subsection*{Combined Torsion and Shear}
\begin{figure} [b]
\centering
\includegraphics[angle=0,scale=0.29]{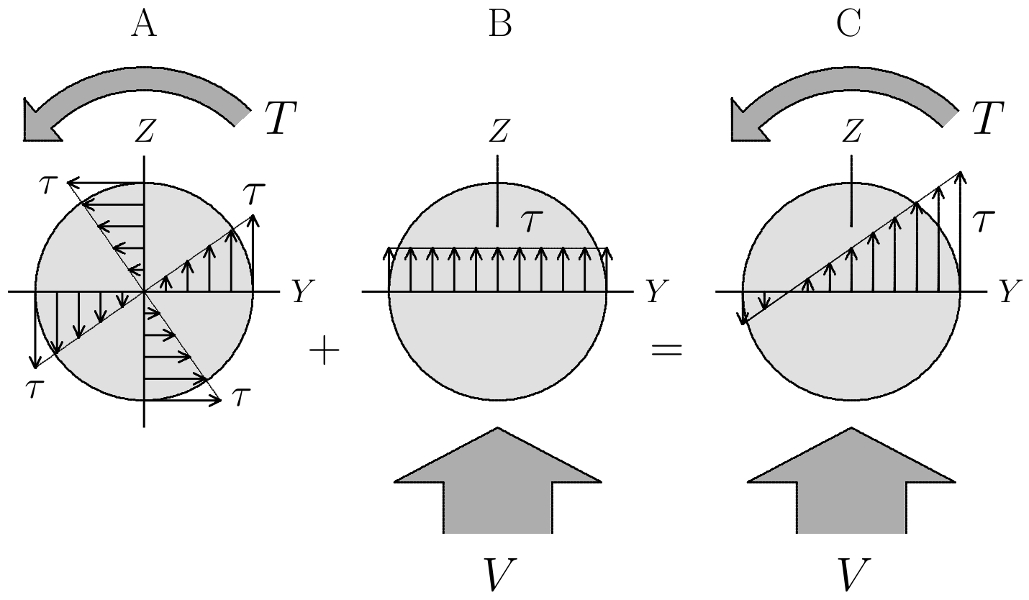}
\caption{Distribution of shear stress on the 
	 cross section of a beam subjected to a transverse load~$V$ 
         in the direction of the positive $Z$-axis and an
         anti-clockwise torque~$T$ about the $X$-axis.
	 In~(A) the radial distribution of stress due exclusively to
         torque is shown along the $Y$- and $Z$-axes, 
         in~(B) the uniform stress due to the vertical force~$V$
         is shown along the $Y$-axis, and in~(C) the superposition 
         of those stresses on the $Y$-axis is shown.
         \label{torsec}}
\end{figure}

We next consider torsion combined with transverse shear. For 
generality and
simplicity of illustration we regard circular cross-sections.
As with bending and axial force, we use 
the superposition principle 
to obtain the stress distribution for the two loads combined.
From the relationship between stress and strain,
generalized Hooke's law, we have
\begin{equation}
    \gamma=\frac{1}{G}\tau
\end{equation}
We further assume for the stress distribution that
\begin{equation}
    \tau=\frac{\rho}{R}\tau_{\rm max}
    \label{TrhoR}
\end{equation}
and for the shear strains that
\begin{equation}
    \gamma=\frac{\rho}{R}\gamma_{\rm max},
    \label{GrhoR}
\end{equation}
where $R$ is the maximum radius of the circular cross 
section and
\begin{equation}
    0<\rho<R
\end{equation}
is the radial distance from the center of the cross section.
Hence stress and strain increase linearly towards the outer
surface where the maximum value is attained for both.
The relationship between applied torque~$T$ and
shear stress on the cross section is
\begin{equation}
    \tau=\frac{T}{J}\rho
    \label{toreq}
\end{equation}
where $J$ is the polar moment of inertia.
We see from Eq.~(\ref{toreq}) that the relationship between~$T$ 
and~$\tau$ is analoguos to the relationship
between~$M$ and~$\sigma$ in Eq.~(\ref{mzi}).
For the average shear due to a vertical force we have
\begin{equation}
    \tau=\frac{V}{A},
    \label{tauva}
\end{equation}
see Fig.~\ref{torsec}B.
Combining Eqs.~(\ref{toreq}) and~(\ref{tauva}), 
\begin{equation}
    \tau=\frac{V}{A}+\frac{T\rho}{J}
    \label{tauva2}
\end{equation}
is the total shear force acting on the cross section.
In Fig.~\ref{torsec}C the two quantities are seen to oppose each other
on the extreme left, while they add up on the extreme right.
Torque is taken to be positive as shown,
i.e., when it is a vector in the positive $X$-direction.

As before, we are not concerned with {\em where} 
on any particular 'beam' the stress is highest. Instead we simply 
identify the maximum stress that occurs with the aim to decide
whether this is above or below the breaking threshold. 
If the element exceeds this threshold then it is removed as
a carrier of force in the elastic equations.
We see that Eq.~(\ref{tauva2}) is analogous 
to~Eq.~(\ref{divThru}) for combined bending and axial force.
We therefore proceed by dividing
through Eq.~(\ref{tauva2}) by the maximum allowable
shear stress~$\tau_{\rm y}$ within the elastic range. We then
obtain
\begin{equation}
    \frac{\tau}{\tau_{\rm y}}=\frac{V}{V_{\rm y}}+\frac{T}{T_{\rm y}},
    \label{divtau}
\end{equation}
where
\begin{equation}
    V_{\rm y}=\tau_{\rm y}A
\end{equation}
is the maximum of the transverse force~$V$ in pure
loading without a torque, and, from Eq.~(\ref{toreq}),
\begin{equation}
    T_{\rm y}=\frac{J}{\rho}\tau_{\rm y}
\end{equation}
is the maximum torque the element can sustain in pure
rotational displacements. Assuming the element fails
when
\begin{equation}
    \tau>\tau_{\rm y}
\end{equation}
our criterion for failure under combined torque 
and shear becomes
\begin{equation}
    \frac{V}{t_{V}}+\frac{T}{t_{T}}>1
\end{equation}
Here we have defined $t_{V}=V_{\rm y}$ and $t_{T}=T_{\rm y}$
as breaking thresholds. 
Requiring only the maximum stress, 
\begin{equation}
    \frac{|V|}{t_{V}}+\frac{|T|}{t_{T}}>1
\end{equation}
is our fracture criterion. 

As for the breaking thresholds we use one threshold for each 
element in our calculations.
Otherwise one might be led to make inferences about the
detailed structure of each 'beam', i.e., such as where 
flaws are located. For instance, 
stress due to applied torque~$T$ is at its greatest furthest away 
from the axis of the element, see Eq.~(\ref{TrhoR}),
while shear stress due to a top-to-bottom vertical 
force~$V$, according to Jourawski's formula, is 
at its greatest across the centre of the section, midway between
the top and bottom surfaces~\cite{carp}. 
Hence, whereas a flaw at the top or bottom surface
will reduce the torque strength substantially, it will 
not to any great extent adversely affect the strength 
with which the element opposes vertical force.
If one chooses to specify different thresholds for the two
terms there are two obvious options. One is to assume a 'realistic' 
distribution of thresholds whereby~$t_{V}$ and~$t_{T}$ are 
correlated so as to take into account different
categories of flaws, the other is to simply assume
that the thresholds are independently random for both types
of loading. 
In our calculations we presently use the same threshold for
both terms. This is based on the notion that 'beam' elements
are the basic building blocks in our system, i.e., they
define the smallest length-scale. All heterogeneity pertaining to
material flaws and/or variations in elastic properties are
assumed to occur on scales at or above that of the individual 
discrete element. 

\subsection*{Combined Axial Force and Shear}
\begin{figure} [t]
\centering
\includegraphics[scale=0.300]{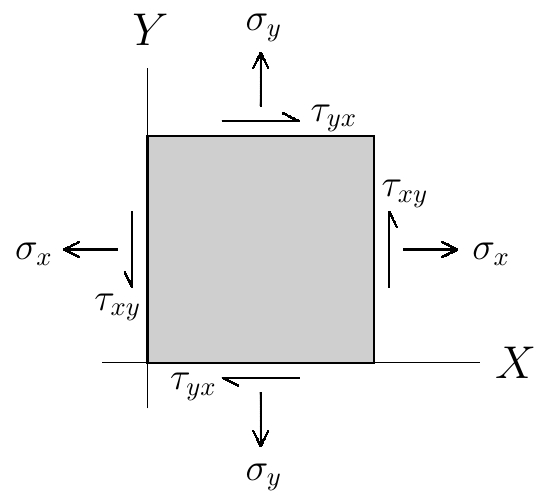}
\caption{Elastic body element showing components of 
         normal stress, $\sigma_{x}$ and~$\sigma_{y}$, and 
         components of shear,~$\tau_{xy}$ 
 	 and~$\tau_{yx}$, on those surfaces that are parallel to
	 the $Z$-axis.
         \label{bodelm}}
\end{figure}

\begin{figure} [b]
\centering
\includegraphics[angle=0,scale=0.32]{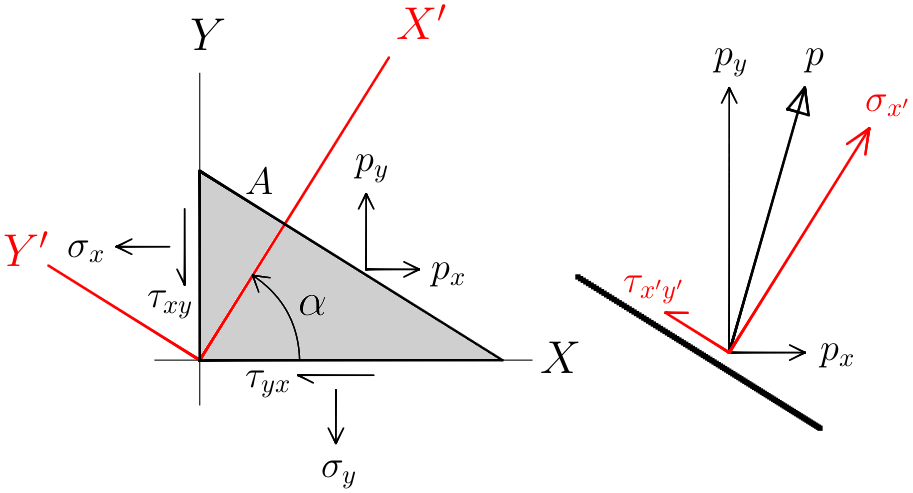}
\caption{Free body with stress components on all
	 surfaces. Decomposition of
	 the stress vector $\bm{p}$ relative to the
         $X^{\prime}Y^{\prime}$-coordinate
	 system is shown in red, 
	 decomposition relative to the $XY$-coordinate
	 system is shown in black.
	 The surface area of the inclined plane is~$A$.
         \label{bodpla}}
\end{figure}

To obtain a general expression for the combined effects of 
axial force and shear we regard a body element under biaxial 
stress, such as that shown in Fig.~\ref{bodelm}.
This body element is in a uniform state of stress.
Stress being a second-order 
tensor, however, stress vectors vary according to the 
surface on which they act. In the following we
regard a plane which intersects the body element at a given
angle and observe
how the components vary as the angle is varied~\cite{chou},
see Fig.~\ref{bodpla}. Here the 
$X^{\prime}Y^{\prime}$-coordinate system has been rotated
through an angle~$\alpha$, such that the
$X^{\prime}$-axis coincides with the normal 
to the inclined plane.
For a body element in equilibrium, the stress vector~$\bm{p}$ 
acting on this plane is obtained by requiring the sum of
forces to be zero. If we resolve the vector $\bm{p}$ in
the~$XY$-coordinate system, we obtain
\begin{equation}
    \bm{p}=\bm{p}_{x}+\bm{p}_{y},
\end{equation}
the components of which are found to be
\begin{equation}
    p_{x}=\sigma_{x}\cos\alpha+\tau_{xy}\sin\alpha
    \label{peex1}
\end{equation}
and
\begin{equation}
    p_{y}=\sigma_{y}\sin\alpha+\tau_{xy}\cos\alpha
    \label{peey1}
\end{equation}
However, $\bm{p}$ can also be resolved in 
the $X^{\prime}Y^{\prime}$-coordinate system. Expressing first
$\sigma_{x^{\prime}}$ and $\tau_{x^{\prime}y^{\prime}}$ in terms
of~$p_{x}$ and~$p_{y}$, as shown on the right
in Fig.~\ref{bodpla}, Eqs.~(\ref{peex1}) and~(\ref{peey1})
are next used to obtain
$\sigma_{x^{\prime}}$, $\sigma_{y^{\prime}}$ and 
$\tau_{x^{\prime}y^{\prime}}$ in terms
of~$\sigma_{x}$, $\sigma_{y}$ and $\tau_{xy}$.
We thus obtain
\begin{equation}
    \sigma_{x^{\prime}}=\frac{\sigma_{x}+\sigma_{y}}{2}
                       +\frac{\sigma_{x}-\sigma_{y}}{2}\cos 2\alpha
		       +\tau_{xy}\sin 2\alpha
    \label{tose1}
\end{equation}
and
\begin{equation}
    \sigma_{y^{\prime}}=\frac{\sigma_{x}+\sigma_{y}}{2}
                       -\frac{\sigma_{x}-\sigma_{y}}{2}\cos 2\alpha
		       -\tau_{xy}\sin 2\alpha
    \label{tose2}
\end{equation}
for the normal stresses, and
\begin{equation}
    \tau_{x^{\prime}y^{\prime}}=\frac{\sigma_{y}-\sigma_{x}}{2}
                                \sin 2\alpha+\tau_{xy}\cos 2\alpha
    \label{tose3}
\end{equation}
for the shear stress in the 
$X^{\prime}Y^{\prime}$-system. 
These are known as the transformation
of stress equations~\cite{chou}, and allow us to determine the stress on any
plane when the angle~$\alpha$ and the stresses~$\sigma_{x}$,
$\sigma_{y}$ and $\tau_{xy}$ are known.

We next seek the extreme values of stress by varying the
orientation of the inclined plane. 
Assuming structural integrity to be
exceeded when the normal stress reaches a critical value, we
evaluate
\begin{equation}
    \frac{{\rm d}\sigma_{x^{\prime}}}{{\rm d}\alpha}=0
\end{equation}
to obtain
\begin{equation}
    \tan 2\alpha=\frac{2\tau_{xy}}{\sigma_{x}-\sigma_{y}}
    \label{taneq1}
\end{equation}
This expression implies two solutions for~$\alpha$ which
are~$90^{\circ}$ apart.
We also see that Eq.~(\ref{taneq1})
is identical to Eq.~(\ref{tose3}), provided that 
$\tau_{x^{\prime}y^{\prime}}=0$. 
Extreme values of normal stress are therefore obtained where 
shear stress vanishes.

Substituting the angles which satisfy Eq.~(\ref{taneq1}) into 
Eq.~(\ref{tose1}) we obtain, after a few 
manipulations,
\begin{equation}
    \sigma_{x^{\prime}}=\frac{\sigma_{x}+\sigma_{y}}{2}
    +\sqrt{\bigl(\frac{\sigma_{x}-\sigma_{y}}{2}\bigr)^{2}
    +\tau_{xy}^{2}}
    \label{sgmx}
\end{equation}
for the maximum normal stress. 'Beam' elements in our
model are force carriers between lattice nodes, hence
there is no normal stress perpendicular to the connecting line
between these and Eq.~(\ref{sgmx}) becomes
\begin{equation}
    \sigma_{\rm m}=\frac{\sigma_{x}}{2}
    +\sqrt{\bigl(\frac{\sigma_{x}}{2}\bigr)^{2}
    +\tau_{xy}^{2}},
    \label{sgmx1}
\end{equation}
where the maximum value of $\sigma_{x^{\prime}}$ has been 
denoted~$\sigma_{\rm m}$. This expression is divided through
by the failure threshold $\sigma_{\rm f}$ for the normal stress
obtained in pure axial loading, to give
\begin{equation}
    \frac{\sigma_{\rm m}}{\sigma_{\rm f}}=\frac{R_{\sigma}}{2}
    +\sqrt{\bigl(\frac{R_{\sigma}}{2}\bigr)^{2}
    +\bigl(kR_{\tau}\bigr)^{2}},
    \label{FFVeq}
\end{equation}
where we have introduced the dimensionless ratios of normal
and shear stress to their respective failure thresholds,
\begin{equation}
    R_{\sigma}=\frac{\sigma_{x}}{\sigma_{\rm f}},
    \qquad
    R_{\tau}=\frac{\tau_{xy}}{\tau_{\rm f}},
    \label{rparam}
\end{equation}
as well as the parameter
\begin{equation}
    k=\frac{\tau_{\rm f}}{\sigma_{\rm f}}
    \label{kparam}
\end{equation}
for the ratio of the shear and normal failure stresses.
This ratio, for steel, is often taken to be in the range~$0.5-0.75$.
For rocks the ratio of tensile to shear strength corresponds
to roughly~$k\sim1$, while the compressive strength is at least
ten times higher than the tensile strength~\cite{afro}.
Assuming that our material fails when
\begin{equation}
    \sigma_{\rm m}>\sigma_{\rm f},
\end{equation}
our criterion for when a 'beam' element fails is
\begin{equation}
    \frac{F}{2t_{F}}
    +\sqrt{\bigl(\frac{F}{2t_{F}}\bigr)^{2}
    +\bigl(k\frac{V}{t_{V}}\bigr)^{2}}>1,
    \label{fcrit1}
\end{equation}
where in Eq.~(\ref{FFVeq}) loads and failure loads have been 
substituted for stresses and failure stresses.

Assuming instead that material integrity is exceeded when
shear stress reaches a critical value, 
\begin{equation}
    \frac{{\rm d}\tau_{x^{\prime}y^{\prime}}}{{\rm d}\alpha}=0
\end{equation}
is evaluated to obtain
\begin{equation}
    \tan 2\alpha=-\frac{\sigma_{x}-\sigma_{y}}{2\tau_{xy}},
    \label{tan2a}
\end{equation}
the right-hand side of which is the negative reciprocal 
of Eq.~(\ref{taneq1}). This implies that the planes of maximum
shear are at an angle of~$45^{\circ}$ with respect to the 
planes of maximum normal stress~\cite{chou}.

From Eq.~(\ref{tan2a}) expressions for~$\cos{2\alpha}$ 
and~$\sin{2\alpha}$ are obtained which are substituted into
Eq.~(\ref{tose3}). This gives
\begin{equation}
    \tau_{x^{\prime}y^{\prime}}=
         \sqrt{
          \bigl(\frac{\sigma_{x}-\sigma_{y}}{2}\bigr)^{2}+
          \tau_{xy}^{2}}
\end{equation}
for the maximum shear stress.
As with Eq.~(\ref{sgmx}) there is no normal stress perpendicular
to the connecting line between nodes and 
\begin{equation}
    \tau_{\rm m}=
         \sqrt{
          \bigl(\frac{\sigma_{x}}{2}\bigr)^{2}+\tau_{xy}^{2}}
	  \label{tauone}
\end{equation}
is obtained by setting~$\sigma_{y}=0$. Assuming the material fails for
\begin{equation}
    \tau_{\rm m}>\tau_{\rm f},
\end{equation}
where $\tau_{\rm f}$ is the failure threshold for the shear stress,
and dividing through Eq.~(\ref{tauone}) by this quantity, we get
\begin{equation}
    \tau_{\rm m}=
    \sqrt{\bigl(\frac{R_{\sigma}}{2k}\bigr)^{2}+R_{\tau}^{2}}
	  \label{tauone1}
\end{equation}
using the first of Eqs.~(\ref{rparam}), and Eq.~(\ref{kparam}).
The criterion for when a 'beam' element should break is then
\begin{equation}
    \sqrt{\bigl(\frac{F}{2k t_{F}}\bigr)^{2}+
          \bigl(\frac{V}{t_{V}}\bigr)^{2}}>1,
	  \label{fcrit2}
\end{equation}
where in Eq.~(\ref{tauone1}) we have substituted loads and failure 
loads for stresses and failure stresses. 

\begin{figure} [t]
\centering
\includegraphics[angle=0,scale=0.30]{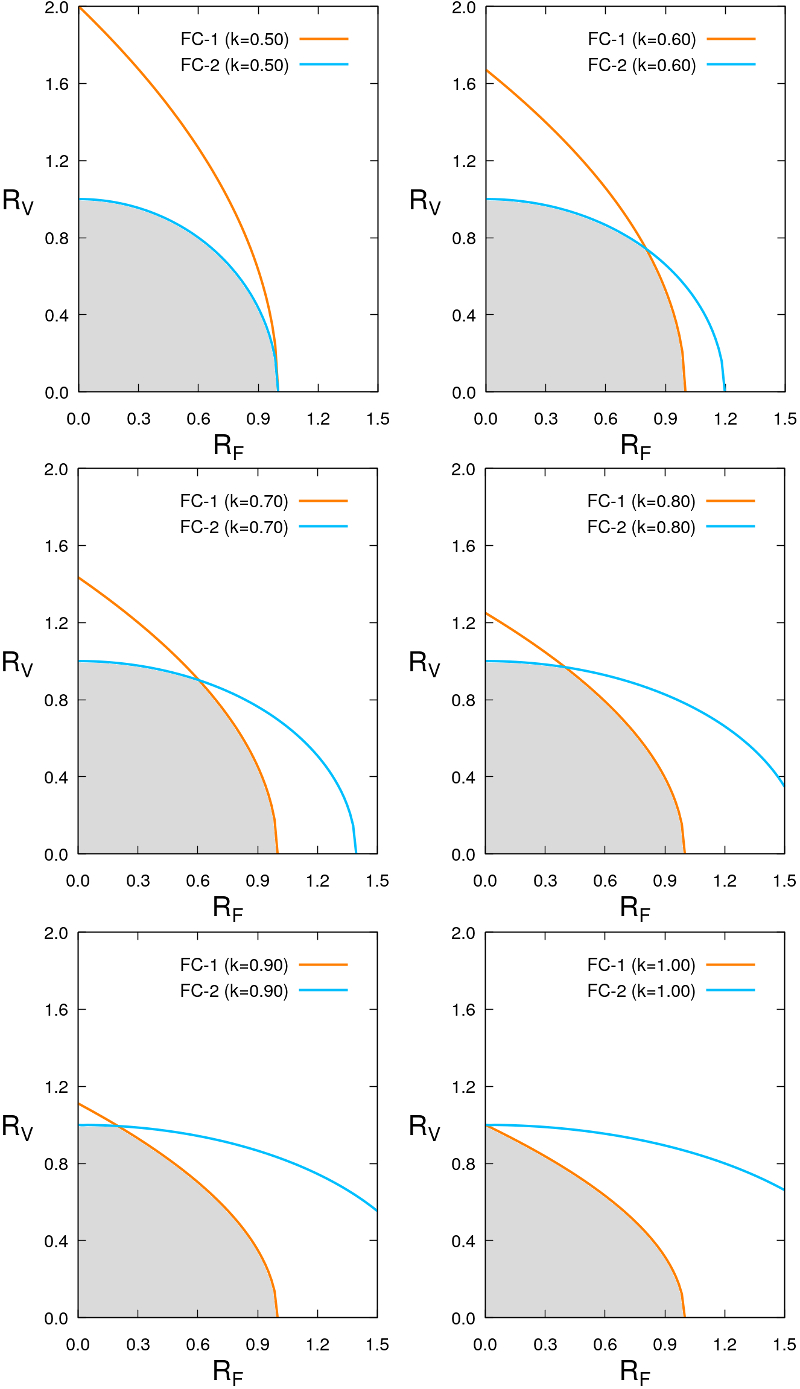}
\caption{Failure envelopes corresponding of Eqs.~(\ref{fcrit1})
	 and~(\ref{fcrit2}), 
	 presently denoted {\rm FC-1} and {\rm FC-2},
	 shown for~$k=0.5$, $k=0.6$, $k=0.7$, $k=0.8$, $k=0.9$
	 and~$k=1.0$. The expressions $R_{F}$ and $R_{V}$
	 are the dimensionless ratios $F/t_{F}$ and $V/t_{V}$,
	 respectively, of the load to failure loads.
         \label{6fck}}
\end{figure}
Plotting Eqs.~(\ref{fcrit1}) and~(\ref{fcrit2}) for different values
of~$k$ it is seen that Eq.~(\ref{fcrit1}) with~$k=1$ 
and Eq.~(\ref{fcrit2}) with~$k=0.5$ provide interaction curves where
the expressions
\begin{equation}
	\frac{F}{t_{F}}<1,\qquad
	\frac{V}{t_{V}}<1
	\label{RfRv}
\end{equation}
are both satisfied, see Fig.~\ref{6fck}.
For values of~$k$ between~$0.5$ and~$1$ the failure envelope is
a combination of Eqs.~(\ref{fcrit1}) and~(\ref{fcrit2}),
corresponding to
the innermost region bounded by the yellow and blue 
curves in Fig.~\ref{6fck} (these have been shaded in gray). 
Eq.~(\ref{fcrit2}) with~$k=0.5$, 
according to Refs.~\cite{nasa} and~\cite{rjwi}, provides a convenient 
and conservative interaction curve when considering combined shear 
and axial loading, this is the shaded region enclosed entirely
by the blue curve in the upper left pane of Fig.~\ref{6fck}. 
Hence, we choose
\begin{equation}
    \sqrt{\bigl(\frac{F}{t_{F}}\bigr)^{2}+
          \bigl(\frac{V}{t_{V}}\bigr)^{2}}>1
	  \label{fcritsh}
\end{equation}
as our criterion.
For $k\rightarrow0.5$ we see from Fig.~\ref{6fck} 
that the shaded region approaches this criterion, 
while it approaches 
\begin{equation}
    \frac{F}{2t_{F}}
    +\sqrt{\bigl(\frac{F}{2t_{F}}\bigr)^{2}
    +\bigl(\frac{V}{t_{V}}\bigr)^{2}}>1
     \label{fcritno}
\end{equation}
when~$k\rightarrow1$ (the yellow curve in the bottom right window). 
From Fig.~\ref{6fck} it is also evident that Eq.~(\ref{fcritsh})
is a good fit within the greater part of the range~$0.5<k<1$,
deviating the most for values above~$k\simeq 0.9$.
For comparison, we will nonetheless also 
include results obtained with Eq.~(\ref{fcritno})
to see if this slightly more conservative alternative makes
any difference. 

\subsection*{Combined Axial Force, Shear, Torsion and Bending}
Finally we seek an interaction formula which combines 
axial force, shear, bending and torsion. 
The basic form of the fracture criterion is taken to be the 
interaction between axial force and shear, as given by
Eq.~(\ref{fcritsh}) or Eq.~(\ref{fcritno}). Within this prescription,
bending is considered in combination with axial force, and
torsion is considered as a contribution to shear.

With Eq.~(\ref{fcritsh}) as our basic expression, the fracture
criterion is then
\begin{equation}
	\sqrt{\bigl(\frac{\widehat{F}}{t}\bigr)^{2}+
	\bigl(\frac{\widehat{V}}{t}\bigr)^{2}}>1
    \label{fcrit02}
\end{equation}
where
\begin{equation}
    \widehat{F}=|F|+|M|
    \label{FMeq}
\end{equation}
is the total stress due to deformations which cause elongation, 
as shown in Fig.~\ref{tensbend}, and
\begin{equation}
    \widehat{V}=|V|+|T|
    \label{VTeq}
\end{equation}
is the total stress from deformations contributing to shear, as 
shown in Fig.~\ref{torsec}. Note that in 
Eq.~(\ref{fcrit02}) we have also assumed the same breaking 
threshold for loading in shear and tension, that is
\begin{equation}
    t=t_{F}=t_{V}
    \label{ttftv}
\end{equation}

In three dimensions the shear force~$V$ in Eq.~(\ref{VTeq}) 
acts within two perpendicular planes. If we consider beam~$1$ 
in Fig.~\ref{6beam}, extending along the positive~$X$-axis, 
shear within the~$XZ$- and~$XY$-planes are combined into 
a bi-planar expression in Eq.~(\ref{VTeq}). Hence, we have
\begin{equation}
    |V|=\sqrt{V_{XY}^{2}+V_{XZ}^{2}},
    \label{VVVxy}
\end{equation}
where $V_{XY}$ and $V_{XZ}$ are the respective contributions 
acting within the two planes. Likewise, in 
Eq.~(\ref{FMeq}) axial force~$F$ is combined with the largest 
of the moments at the ends of the 'beam' element, i.e.,
\begin{equation}
    |M|={\rm max}\bigl(M_{i},M_{j}\bigr)
    \label{maxM}
\end{equation}
where $i$ is the near (node) end and $j$ is the far (neighbouring
node) end of the 'beam' element. If we again consider beam~1
we now have
\begin{equation}
	M_{i}=\sqrt{M_{y,i}^{2}+M_{z,i}^{2}},
    \label{maxcM}
\end{equation}
with $M_{y,i}$ and~$M_{z,i}$ representing the contributions from
bending obtained within the $XZ$- and $XY$-planes, respectively 
($M_{y}$ is the bending moment about the cross-sectional centroidal 
axis~$Y$). The expression for $M_{j}$ is similar. 

Finally, for the sake of comparison, the $k=1$ criterion based on 
maximum normal stress is also included. Hence,
Eq.~(\ref{fcritno}) reads
\begin{equation}
    \frac{\widehat{F}}{2t}
    +\sqrt{\bigl(\frac{\widehat{F}}{2t}\bigr)^{2}
    +\bigl(\frac{\widehat{V}}{t}\bigr)^{2}}>1,
     \label{fcrit01}
\end{equation}
where the quantities~$\widehat{F}$,~$\widehat{V}$ and~$t$ are
given by the same expressions as those in~Eq.~(\ref{fcrit02})
above.

Although a 'beam' element, when regarded as a separate entity, 
can be strained, twisted and deformed in all manner of ways,
we regard independent couplings between torsion and 
bending as less significant.
Interaction between these two effects is still included, but only
indirectly in the sense that
bending contributes to axial stress, and torsion to shear,
{\em before} the two are combined via Eqs.(\ref{fcritsh})
or~(\ref{fcritno}).
This also applies to the combination of shear and bending, and to
the combination of axial deformation and torsion -- any direct 
interaction between these effects is assumed negligible.

Our assumption of a fracture criterion taking this form is not
unreasonable considering the fact that we intend to model a 
continuum, within which the 'beam' is embedded and thereby 
considerably constrained by the surrounding medium.  
The situation would be different in considering an isolated 
beam which can move freely, and even more so if this beam is 
of the slender type or has a cross-sectional geometry that is 
important in the overall context. 
Moreover, in modeling a discretized
continuum, realistic forces between nodes should preclude the
use of discrete elements based on slender beams.

\section{Illustration of Stresses in Combined Loads}
\label{illstress}
In order to substantiate how stresses are distributed throughout a 
structure when combined loadings are applied we
can construct 'macroscopic' beams from discrete elements. Based
on a cubic lattice morphology, the simplest such structure is
a square prism. Structures with other cross-sections
are obtained by 
\begin{figure} [t]
\centering
\includegraphics[scale=0.30]{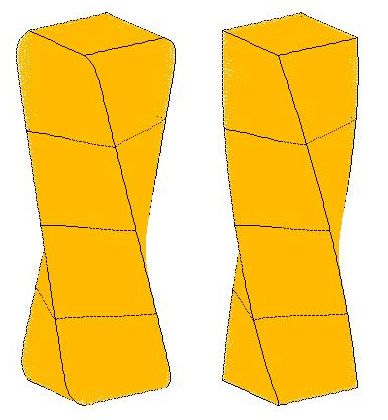}
\caption{A square prism beam deformed in torsion. 
	 The top surface has been rotated $45^{\circ}$ clockwise 
	 and the bottom surface $45^{\circ}$ counter-clockwise.
	 Version based 
	 on linear equations is shown on the left, and version 
	 based on non-linear equations is shown on the right.
         \label{twistwarp}}
\end{figure}
cutting away beams that lie outside the required geometric
profile. Circular or elliptic 
cross-sections, for instance, are easily obtained in this way.
Presently we regard beams with square cross-sections and marked 
outlines, since it is easier to visualize deformations 
(especially torsion) this way.

Deformations are best visualized when they are sizeable. The 
displacements involved in our calculations for the roughness exponent,
on the other hand, are quite small. This is appropriate in a
brittle fracture study, since the external displacements involved
are usually small. Large deformations are more commonly 
associated with ductile materials. However, in order to illustrate
a few cases of how stresses are altered as different modes of
external loading are combined, we employ large displacements
for visual effect. 
If we use Eqs.~(\ref{forceXi}) to~(\ref{forceWi})
for this purpose a 'warping' effect is obtained when 
angular displacements become large. This is shown on the
left in Fig.~\ref{twistwarp}, where the top and bottom surfaces
have been rotated in opposite directions.
Edges near the top and bottom are seen to turn 
inwards after a gradual swelling develops as the ends are approached. 
The reason why the effect is most marked near the ends is 
because it is here that rotational displacements are largest. 

\begin{figure} [b]
\centering
\includegraphics[scale=0.38]{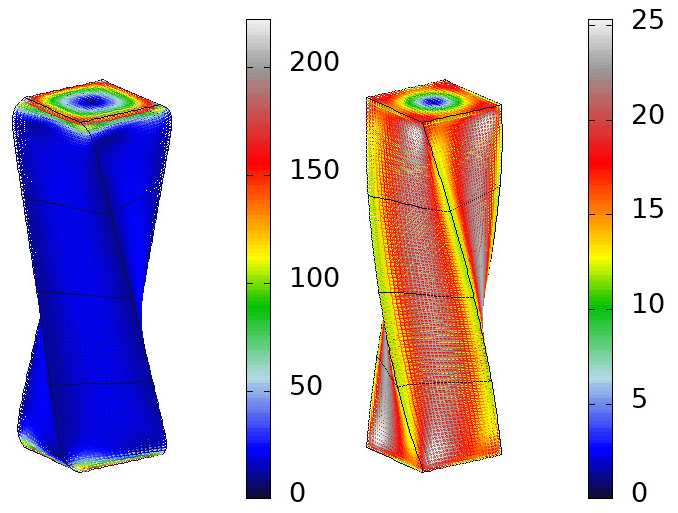}
\caption{Shear stresses in square beam with large torsionl
	 deformations. Version based 
	 on linear equations is shown on the left, and version 
	 based on non-linear equations is shown on the right.
         \label{2torx}}
\end{figure}
In contrast to this is the version shown on the right in 
Fig.~\ref{twistwarp}. Here, the axial 
contributions from the beams have been corrected to take into account 
the rotations about three axes. Components of shear and 
bending moment should also be adjusted in this way, leading to 
equations which involve a large number of terms. However, provided 
deformations are not too extreme, corrections 
applied to the axial terms are the most important. In 
Fig.~\ref{twistwarp} the rotation of the top and bottom surfaces is 
45 degrees, for a total rotation of 90 degrees between top and bottom. 
For such a large deformation the version on the right represents 
a dramatic improvement 
over the one on the left. The relevant modifications to 
Eqs.~(\ref{forceXi}),~(\ref{forceYi}) and~(\ref{forceZi}) are 
\begin{widetext}
\begin{eqnarray}
	X_{i}=
	  &&\hspace{-6mm}
	-\frac{1}{\alpha}
	  \sum_{j=1}^{6}
	        \widehat{\lambda}_{1,3}
		\bigl(1-\sqrt{A_{j}}\hspace{0.5mm}\bigr)s_{j}\Phi_{j}
	  -\frac{1}{\beta\hspace{-0.5mm}+\hspace{-0.5mm}\frac{\gamma}{12}}
	     \sum_{j=1}^{6}
	     \widehat{\chi}_{1,3}
	     \biggl\{\delta x_{j}
             +\frac{r_{j}}{2}
         \Bigl[\widehat{\lambda}_{5,6}\bigl(v_{i}+v_{j}\bigr)+
               \widehat{\lambda}_{2,4}\bigl(w_{i}+w_{j}\bigr)
         \Bigr]\biggr\},
		\label{modXi}
\end{eqnarray}
for the $X$-component of force on node~$i$,
\begin{eqnarray}
	Y_{i}=
	  &&\hspace{-6mm}
	-\frac{1}{\alpha}
	  \sum_{j=1}^{6}
	        \widehat{\lambda}_{2,4}
		\bigl(1-\sqrt{A_{j}}\hspace{0.5mm}\bigr)s_{j}\Phi_{j}
	  -\frac{1}{\beta\hspace{-0.5mm}+\hspace{-0.5mm}\frac{\gamma}{12}}
	     \sum_{j=1}^{6}
	     \widehat{\chi}_{2,4}
	     \biggl\{\delta y_{j}
             +\frac{r_{j}}{2}
         \Bigl[\widehat{\lambda}_{5,6}\bigl(u_{i}+u_{j}\bigr)+
               \widehat{\lambda}_{1,3}\bigl(w_{i}+w_{j}\bigr)
         \Bigr]\biggr\}
		\label{modYi}
\end{eqnarray}
for the $Y$-component, and
\begin{eqnarray}
	Z_{i}=
	  &&\hspace{-6mm}
	-\frac{1}{\alpha}
	  \sum_{j=1}^{6}
	        \widehat{\lambda}_{5,6}
		\bigl(1-\sqrt{A_{j}}\hspace{0.5mm}\bigr)r_{j}\Phi_{j}
	  -\frac{1}{\beta\hspace{-0.5mm}+\hspace{-0.5mm}\frac{\gamma}{12}}
	     \sum_{j=1}^{6}
	     \widehat{\chi}_{5,6}
	     \biggl\{\delta z_{j}
             -\frac{s_{j}}{2}
         \Bigl[
               \widehat{\lambda}_{2,4}\bigl(u_{i}+u_{j}\bigr)+
               \widehat{\lambda}_{1,3}\bigl(v_{i}+v_{j}\bigr)
         \Bigr]\biggr\},
		\label{modZi}
\end{eqnarray}
for the $Z$-component. Here
\begin{eqnarray}
	A_{j}
	  \hspace{-2mm}
	 &=&
          \hspace{-1mm}
	  \widehat{\chi}_{1,3}\hspace{0.5mm}\delta x_{j}^{2}
	 +\widehat{\lambda}_{1,3}\bigl(1-r_{j}\delta x_{j}\bigr)^{2}
         +\hspace{-1mm}
	  \widehat{\chi}_{2,4}\hspace{0.5mm}\delta y_{j}^{2}
         +\widehat{\lambda}_{2,4}\bigl(1+r_{j}\delta y_{j}\bigr)^{2}
	 +\hspace{-1mm}
	  \widehat{\chi}_{5,6}\hspace{0.5mm}\delta z_{j}^{2}
         +\widehat{\lambda}_{5,6}\bigl(1+r_{j}\delta z_{j}\bigr)^{2}
	 \nonumber
\end{eqnarray}
is the squared length of the discrete element between nodes~$i$ 
and~$j$ (disregarding curvature). Furthermore,
\begin{equation}
	\Phi_{j}=\widehat{\lambda}_{1,3}\cos{v_{i}}\cos{w_{i}}
	        +\widehat{\lambda}_{2,4}\cos{u_{i}}\cos{w_{i}}
	        +\widehat{\lambda}_{5,6}\cos{u_{i}}\cos{v_{i}}
\end{equation}
is the angular displacement of node~$i$. 
As can be seen from Fig.~\ref{twistwarp}, the version
on the right is clearly free of the warping seen in the 
version on the left.
\end{widetext}

The importance of taking into account local rotations for
large deformations is made even more clear if we regard the
stresses involved. In Fig.~\ref{2torx} the shear stresses
involved in the two cases are shown, and the colour scales included 
with the beams illustrate the point. Evidently, when using linear 
equations, the
stresses involved near the ends of the beam become quite extreme,
almost ten times higher than elsewhere in the beam. Away from
the ends, however, the stresses are comparable, as can be seen 
from the colour scales. In contrast, a uniform distribution of 
shear is obtained with Eqs.~(\ref{modXi}) to~(\ref{modZi}) 
in place of Eqs.~(\ref{forceXi}) to~(\ref{forceZi}).

\begin{figure} [b]
\centering
\includegraphics[scale=0.35]{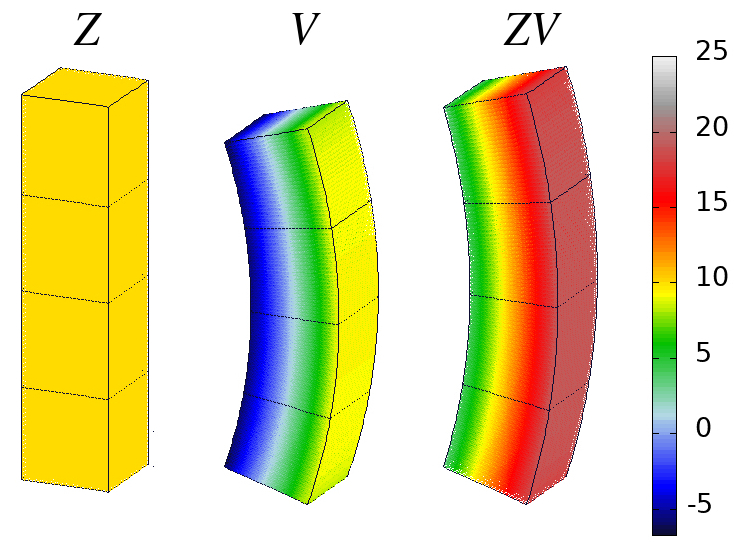}
\caption{A square prism beam structure, sides 
	 $L\hspace{-0.5mm}=\hspace{-0.5mm}25$ and
	 height $H\hspace{-0.5mm}=\hspace{-0.5mm}101$, loaded in 
	 tension ({\it Z\/}), pure bending ({\it V}\/), and the two 
	 combined ({\it ZV\/}).
         \label{axibeamZV2}}
\end{figure}
The relevant equations are more complicated, and involve non-linear
terms which necessitate an iterative adaption of
conjugate gradients. In this approach 
the decreasing residuals of each successive solution are adopted 
as a starting point for a new tentative solution.  Hence, at each 
stage in the breaking process a loop produces a sequence of tentative 
solutions. Within this loop, the number of iterations required for 
each successive solution decreases rapidly until the solution has 
converged. Although computational time increases significantly in 
comparison with the linear set of equations, it is still a only 
a matter of a minute or two to obtain stress distributions for 
relatively large structures. Such structures may be intact or at 
a pre-determined stage of breaking. 
However, for the purpose of studying the {\em entire} fracture
process it is more practical to use linear equations in
conjunction with small deformations.

We consider a beam in the form of a square prism, where all edges
have been drawn black as an aid to emphasize body shape and 
displacements. External 
loads are imposed by rotating or
translating the top and bottom surfaces of the 
body. The combination
of bending and axial tension discussed in Section~\ref{fcrit} is
illustrated in Fig.~\ref{axibeamZV2} in the case of a beam with 
length~$4L$. 
Additional black lines in the figure delineate cubes 
with sides~$L=25$. On the left is a beam that is loaded
in the vertical direction (stretched along the $Z$-axis), 
in the middle is the same beam in a bent configuration only
(bending within the $XZ$-plane), and on the right the two loadings
are combined. Stresses shown are axial stresses in 
discrete elements
aligned along the vertical axis (the $Z$-axis). 
Referring to the colour scale (the same scale relates to all
three loadings) axial stresses of the tensile and bending 
cases are clearly seen to be additive, as expected from
the superposition of forces in
Eq.~(\ref{FMM2}).

\begin{figure} [t]
\centering
\includegraphics[scale=0.35]{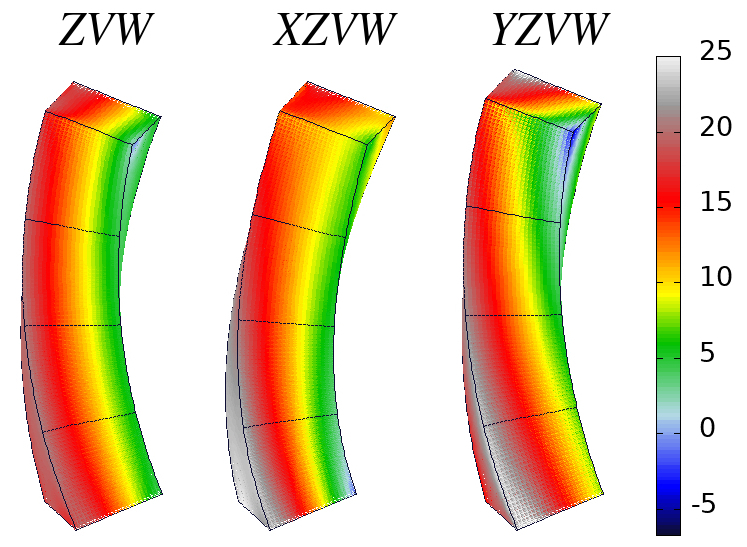}
\caption{A square prism beam, with sides 
	 $L\hspace{-0.5mm}=\hspace{-0.5mm}25$ and
	 height $H\hspace{-0.5mm}=\hspace{-0.5mm}101$, loaded in 
	 tension and bending ({\it ZV\/}), with added torsion ({\it ZVW\/}) and
	 shear ({\it XZVW\/} and {\it YZVW\/}).
         \label{combYZVW}}
\end{figure}
\begin{table*}
\caption{\label{ttable} 
Average forces summed from top to bottom along the middle of 
the vertical sides of a structure with square cross-section,
sides $L=25$ and height $H=101$, with the structure having been 
subjected to different external loadings. Loading types denoted
$X$, $Y$ and $Z$ represent translations of the upper
surface along said positive axes. Size of the translations
in all cases corresponds to 10 discrete element lengths. 
Loadings denoted $U$, $V$ and $W$ are rotations about 
axes parallel with the $X$-, $Y$-, and $Z$-axes, respectively.
In all such cases the bottom surface has been rotated $+0.1\pi$ while
the top surface has been rotated $-0.1\pi$. The side facing the viewer
is denoted '000', other sides being '090'~(right), 
'180'~(opposite) and '270'~(left).}
\begin{ruledtabular}
\begin{tabular}{ccrrcccrrc}
& \multicolumn{3}{c}{Axial average\footnote{Average along vertical 
                     elements.}}
		     &&&
\multicolumn{3}{c}{Shear average\footnote{Using Eq.~(\ref{VVVxy}) 
		   to combine shear in the {\it XZ}- 
                   and {\it YZ}-planes.}} &
\\
& Loading 
& $\sum_{Z}F_{090}$ 
& $\sum_{Z}F_{270}$ 
&&&
Loading 
& $\sum_{Z}V_{000}$ 
& $\sum_{Z}V_{180}$ 
&
\\ 
\hline
& {\it Z}     & 10.00 & 10.00 &&& {\it X}          & 5.07 &  5.07 & \\
& {\it V}     &  9.15 & -5.93 &&& {\it W}          & 9.67 &  9.67 & \\
& &&&&& \\		
& Eq.~(\ref{FMeq}) & 19.15 &  4.07 &&& Eq.~(\ref{VTeq}) & 4.60 & 14.74 & \\
& &&&&& \\		
& {\it ZV}    & 19.00 &  3.93 &&& {\it XW}         & 4.76 & 14.57 & \\
& {\it ZVW}   & 19.49 &  4.03 &&& {\it XWV}        & 6.42 & 12.96 & \\
& {\it YZVW}  & 20.52 &  3.78 &&& {\it XZWU}       & 6.25 & 13.14 & \\
& {\it XZVW}  & 19.92 &  4.46 &&& {\it XWUV}       & 4.77 & 14.60 & \\
\end{tabular}
\end{ruledtabular}
\end{table*}
In the figure, positive values are tensile while negative 
values are compressive. 
Average axial forces, 
obtained by summing from top to bottom along 
the middle of the vertical faces of the beam are included
in Table~\ref{ttable}. Here, $\sum F_{090}$ 
and $\sum F_{270}$ 
refer to the convex and concave sides of the beam, respectively, 
in Fig.~\ref{axibeamZV2} 
The reason we consider average values of axial force is because
we presently regard the $25\times25\times101$ square prism 
as a model of a discrete element 'beam'. In a fracture criterion
we will not be interested in details pertaining to scales smaller
than that of each element.

Table~\ref{ttable} shows that
values obtained by adding '$Z$' and~'$V$', as
dictated by Eq.~(\ref{FMeq}), agree well with the actual values
obtained in the combined loading, 
denoted~'$ZV$' in Fig.~\ref{axibeamZV2}\footnote{The discrepancy
is mostly due to the fact that the small vertical shrinkage
required in a pure bending in case '$V$' has not been taken into
account. This would be necessary for the arc-length in the
middle of the beam to equal the length of the beam in its relaxed
state. The remaining discrepancy is due to the moment, being
applied at the top and bottom ends, not being
constant throughout the length of what is essentially a thick
beam.}.
To what extent will additional loadings alter this picture?
Instead of carrying out a systematic investigation involving
many data points, a few
extra loads 
added onto the '{\it ZV}' combinations have been included
in the last three lines of Table~\ref{ttable}. These are
torsion, '{\it ZVW}', as well as torsion and shear, '{\it YZVW}' 
and '{\it XZVW}'. In the latter cases the top surface has been
translated along the positive $Y$- and $X$-axes, respectively. 
Although the displacements and
rotations involving all the loadings are quite sizeable the
average axial forces on the convex ($\sum F_{090}$) and concave
($\sum F_{270}$) sides do not change much, relatively speaking, as
can be seen from the values in Table~\ref{ttable}. Hence, the
task of identifying the largest contribution from axial forces
seems to be adequately taken care of by the first term, 
$\widehat{F}/t$, in Eq.~(\ref{fcrit02}). Had we considered the
detailed distribution of forces rather than the averages, the 
highest axial force in the '{\it YZVW}' and '{\it XZVW}' cases
would have been found to occur in discrete elements that are
situated in the corners of the
cross-section, near the top and bottom surfaces, on the concave
side of the structure (see the colour scale in Fig.~\ref{combYZVW}). 
Numerical
values in these two cases are about 20\% higher than the averages 
quoted in Table~\ref{ttable}. Maximum axial stress in the
'{\it ZVW}' case is only about 5\% higher and occurs in a corner
about three quarters of the way towards the top surface.
In this context we also have to keep in mind that these values
refer to cross-sections that are square rather than circular. 

\begin{figure*} [t]
\centering
\includegraphics[angle=0,scale=0.40]{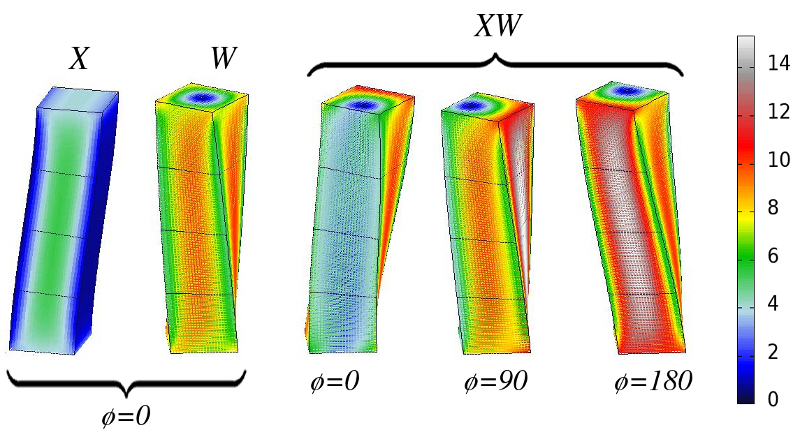}
\caption{{\it Beam of square cross-section, 
	      sides $L=25$ and 
	      height $H=101$, made up of discrete elements. 
	      The structure has been
	      subjected to shear~({\it X\/}), 
	      torsion~({\it W\/}) and both these 
	      loadings combined ({\it XW\/}). 
	      Here, $\phi\ne0$ is a
	      counter-clockwise rotation of the structure. Hence
	      $\phi=90$ has turned the side corresponding to
	      $\sum_{Z}V_{090}$ towards the viewer.
         }
         \label{shebeamXW}}
\end{figure*}
Table~\ref{ttable} also includes average shear forces calculated
along the vertical faces of the square prism structure.
Hence, the combination of shear and torsion, and to what extent this
is affected by other loading modes, is investigated next.
In Fig.~\ref{shebeamXW} is shown a beam under pure shear, 
pure torsion and the combination of these two loadings. The
combined loading has been shown from three different angles,
illustrating how shear on one side increases while that on
the other side decreases, in accordance with the superposition
of forces in Eq.~(\ref{VTeq}).

\begin{figure} [b]
\centering
\includegraphics[scale=0.34]{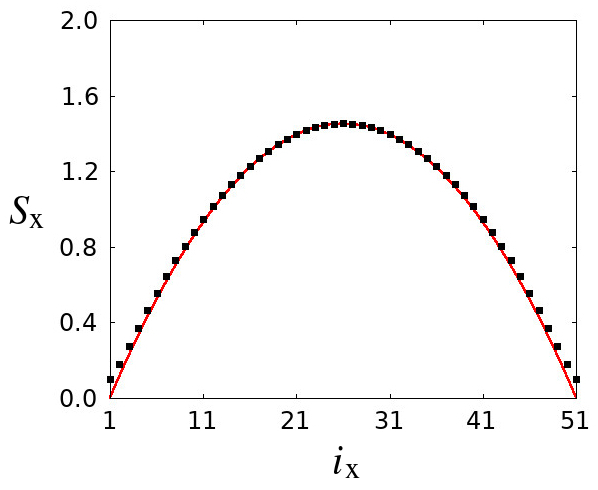}
\caption{Shear force variation across the width of the 
	 mid-section of a $51\times51\times409$ square beam, 
	 with parabola shown in red.
         \label{jourawski51}}
\end{figure}

On the extreme left is a beam where the top surface has been translated
along the positive $X$-axis. The shear force is seen to be at its
largest in the middle of the cross-section and decreases to 
zero at the edges. This is an example of Jourawski's formula, i.e.,
\begin{equation}
    \tau=\frac{V_{x}Q_{y}}{I_{y}L},
    \label{jourawski}
\end{equation}
where
\begin{equation}
    Q_{y}=\int_{A}x{\rm d}A
\end{equation}
is the first, or static, moment of area about the axis
of bending (the $Y$-axis in this case), $I_{y}$ is the moment 
of inertia about the same axis, and $L$ is the width of the 
cross-section. 
For a square cross-section the distribution of forces across 
the width is in the shape of a parabola. This is so because the external 
transverse force~$V_{x}$ produces a bending moment which varies 
along the length of the structure~\cite{carp}, i.e., the $Z$-axis 
in this case. The distribution of shear forces is shown in 
\begin{figure} [b]
\centering
\includegraphics[scale=0.34]{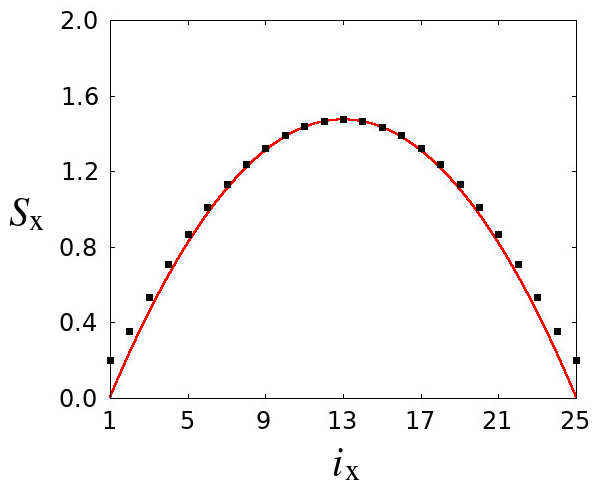}
\caption{Shear force variation across the width of the
	 mid-section of a $25\times25\times201$ square beam, 
	 with parabola shown in red.
         \label{jourawski25}}
\end{figure}
Fig.~\ref{jourawski25}
for a structure which has sides~$L=25$ and vertical length (or height) $8L$. 
The external force $V_{x}$ arises from a horizontal translation of
the top surface a distance of 15 discrete elements.
Numerical values are shown as solid squares
and have been obtained along a line through the centre of the cross-section,
midway between the top and bottom surfaces. 
A parabola, shown in red, has been fit to these values. Apart from
at the very edges, the numerical values are seen to conform very well
to the shape predicted by Jourawski's formula. The discrepancy at the edges
are finite-size effects, as can be seen by making finer the resolution
of the discretization. Increasing proportionally
the dimensions of the sample and the magnitude of the external 
deformation, the effect obtained is analogous to 
such a refinement of numerical resolution.
One example of this is included in Fig.~\ref{jourawski51}, which 
shows~the distribution of shear forces in a structure with sides $L=51$ and
height $8L$. Here the external 
transverse displacement used is 30 discrete elements, 
and the discrepancy at the edges between the numerical 
values and the parabola is now seen to be much smaller.

\begin{figure} [t]
\centering
\includegraphics[scale=0.29]{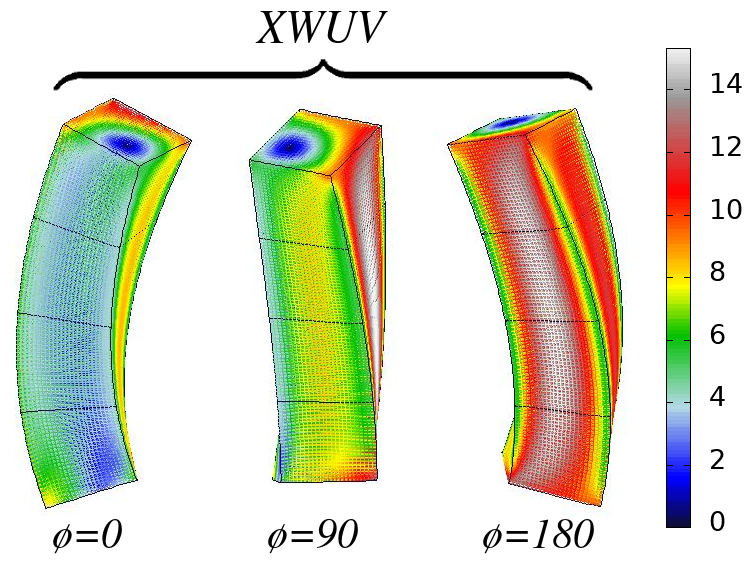}
\caption{A square prism beam, with sides 
	 $L\hspace{-0.5mm}=\hspace{-0.5mm}25$ and
	 height $H\hspace{-0.5mm}=\hspace{-0.5mm}101$, subject to 
	 shear ({\it X\/}), torsion ({\it W\/}) and biaxial 
	 bending ({\it U\/} and {\it V\/}).
         \label{shebeamXWUV}}
\end{figure}

The next beam in Fig.~\ref{shebeamXW}, denoted '{\it W}', is 
under pure torsion, and following this is the '{\it XW}' case 
where the two loadings '{\it X}' and '{\it W}' are combined. 
As expected, values in Table~\ref{ttable} obtained
for the average forces on the sides where shear increases or
decreases compare favourably with values expected from 
Eq.~(\ref{VTeq}), i.e., shear intensifies on one side and is 
alleviated on the other. The interesting question is to what
extent other deformations influence this relationship. Adding
a bending moment '{\it V}' or a biaxial bending 
moment '{\it UV}' changes the values somewhat (see Table~\ref{ttable}, 
but not anywhere
near what would be required to invalidate
the relationship
between '{\it X}' and~'{\it W}' as incorporated 
into Eq.~(\ref{fcrit02}) by the term $\widehat{V}/t$.
The distribution of forces involved when adding the 
biaxial moment~'{\it UV}'
are shown in Fig.~\ref{shebeamXWUV}.

\subsection*{Breaking Thresholds}
The breaking thresholds for the two terms in Eq.~(\ref{fcrit02})
have been set to the same value in Eq.~(\ref{ttftv})
since we wish to avoid making inferences
about the detailed structure of each 'beam'. 
If a 'beam' is axially weak we also assume that it will be weak in
shear, bending and torsion. 
We will not devise
individual threshold distributions based on where 'flaws' in
a 'beam' might be located. Otherwise one might, for instance,
expect a 'beam' with an edge crack to be unaffected in strength when 
bent such that the edge crack is compressed (closed) while weakened 
when it is bent the other way. Likewise, a 'beam' with a central flaw
might be expected to show structural resilience towards bending
in either plane while being weakend in axial strength.
We may still adjust the thresholds so that the 'beam' is proportionally 
weaker in tension than in shear. This can be done, for instance, 
by multiplication with a constant factor. The main point is that 
all thresholds relevant to any given 'beam' is chosen from 
a single value in the stochastic distribution.

\section{External Uniaxial Tension on a Cube}
\label{unicube}
Using the model described in Section~\ref{secmodel}, 
tensile fracture is initiated by imposing
a uniform displacement vertically (along the $Z$-axis) on the top
surface of the lattice. The edges of the cube are taken to be
parallel with the coordinate axes. Discrete elements are removed
one at a time, and at any stage in the fracturing process
Eq.~(\ref{ma3x}) is used to calculate new displacements 
after a discrete element has been removed. The resulting distribution of
stress, in conjunction with the breaking thresholds assigned, 
is used to identify which discrete element will break next. 
Exactly how this identification is made relies on the nature 
of the fracture criterion.

\begin{figure} [b]
\centering
\includegraphics[angle=0,scale=0.35]{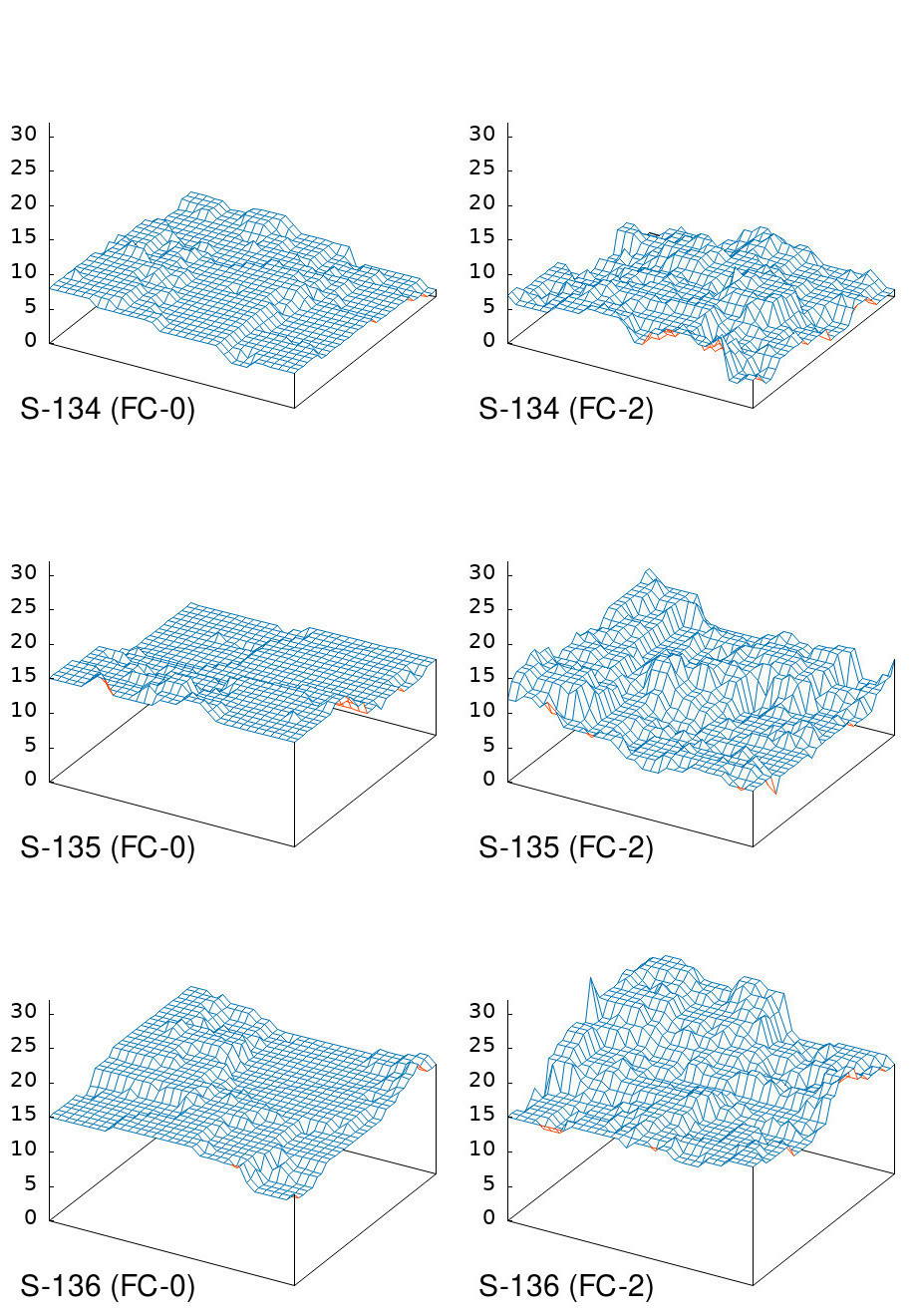}
\caption{Comparison of fracture surfaces obtained with different
	 fracture criteria. 
	 Three different samples are shown, {\rm S-134, S-135 
	 and S-136}. 
	 {\rm FC-0} denotes the 'original' criterion, 
	 Eq.~(\ref{f2m}), and {\rm FC-2} denotes the 'maximum
	 shear stress' criterion, Eq.~(\ref{fcrit02}).
	 Fracture interfaces are red on the underside and
	 blue on top.
         \label{3x2cube_134}}
\end{figure}

Fracture surfaces obtained for three different samples of 
size~$L=32$ are shown in Fig.~\ref{3x2cube_134}. 
The disorder used is one of intermediate strength, corresponding 
to~$D=1$ in the prescription outlined in Section~\ref{disrd}. 
For each sample the only difference between the one on the left and 
its counterpart on the right is the fracture criterion used. Samples
on the left have been broken with the original fracture criterion,
Eq.(\ref{f2m}), while Eq.(\ref{fcrit02}) has 
been used for those on the right.
For the three samples shown the fracture
surfaces appear roughly at the same position vertically on the 
lattice. Slopes, elevated areas and depressions sometimes also 
appear in the same locations. Although
some samples appear superficially
similar for the two fracture criteria, others again
differ substantially. 
A closer look at the three samples in Fig.~\ref{3x2cube_134}, 
however, reveals an important difference between
fracture surfaces obtained with these two criteria. 
This difference, moreover, pertains to {\em all} samples.
Specifically, those obtained
with~Eq.~(\ref{fcrit02}) display a pronounced roughness, in 
stark contrast with those obtained with~Eq.~(\ref{f2m}).
In the latter case fracture
surfaces are seen to consist of flat sections that 
are stepped up or down relative to each other -- reminiscent of a
landscape of 'plateaus'. Fracture surfaces 
evidently look very different depending on which of the two
criteria one uses.

\begin{figure} [b]
\centering
\includegraphics[angle=0,scale=0.50]{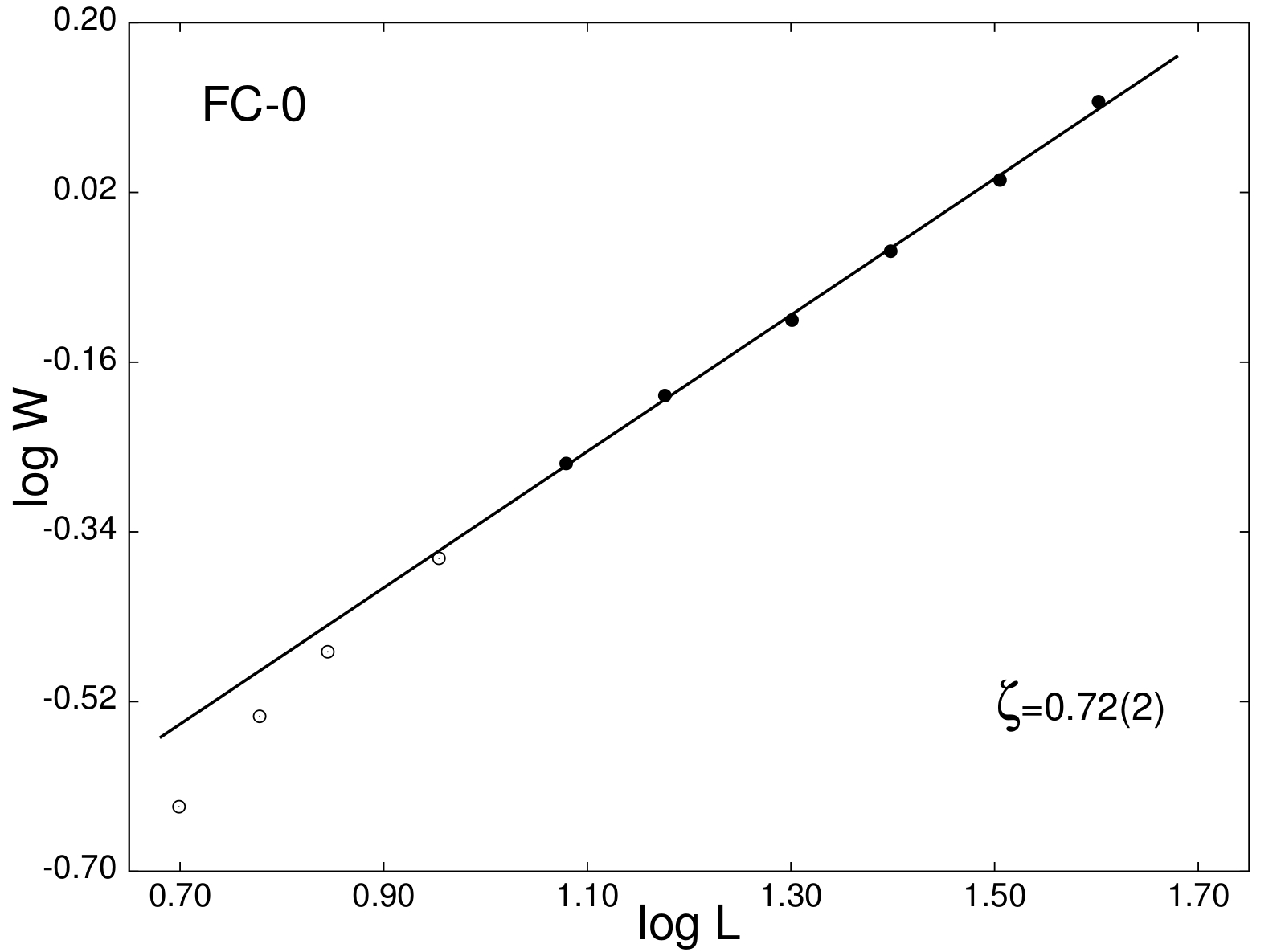}
\caption{Log-log plot showing average roughness, $W$, as a function
         of system size, $L$, for a large number of
	 fractured samples of each size. 
	 Disorder magnitude is $D=1.5$. 
	 {\rm (FC-0)} denotes the result obtained with 
	 the 'original' fracture criterion, Eq.~(\ref{f2m}).
         \label{d1p5_fc0}}
\end{figure}
Such a difference in appearance could, however, also be obtained with
the {\em same} fracture criterion by increasing or decreasing the 
magnitude of structural disorder~\cite{skj2}.
The question is: does the morphology 
change in more fundamental ways than just to provide an
offset in the roughness with respect to disorder strength?
To answer this we turn to a standard yardstick in brittle fracture 
calculations, i.e., the exponent which characterizes how surface 
roughness scales with system size~\cite{mand,darc,smod,adva}. 
Fracture surfaces have been found to be self-affine, meaning that 
if lengths within the fracture plane are scaled by a factor $\lambda$
then lengths perpendicular to this plane scale by a 
factor $\lambda^{\zeta}$, where $\zeta$ is the roughness exponent.
A self-affine relationship $W\sim L^{\zeta}$ is therefore obtained,
and this appears as a straight line in a log-log plot.
Results in 3D have been obtained for $\zeta$ with various models,
such as the random fuse model~\cite{batr,rais,alav,nuka}, which
is an electrical analogue to fracture, and with networks of
elastic springs~\cite{pari}. Results have also been obtained with the
beam lattice~\cite{skj2,bara} with results varying according to 
the fracture criterion used, and possibly also with other parameters 
involved, such as disorder.

Quantification of surface roughness is done
in the same way as in Ref.~\cite{skj2},
i.e., as the root-mean-square variance perpendicular to 
the fracture plane,
\begin{eqnarray}
    W_{x}(L)=\left\langle\frac{1}{L}\sum_{i=1}^{L}z_{x}(i)^{2}-
          \biggl[\frac{1}{L}\sum_{i=1}^{L}z_{x}(i)\biggr]^{2}
           \right\rangle^{1/2},
\end{eqnarray}
where $z_{x}(i)$ is the vertical height
of the first intact node encountered when moving down
towards the lower remaining part of the structure (shown
in Fig.~\ref{3x2cube_134}).

\begin{figure} [t]
\centering
\includegraphics[angle=0,scale=0.35]{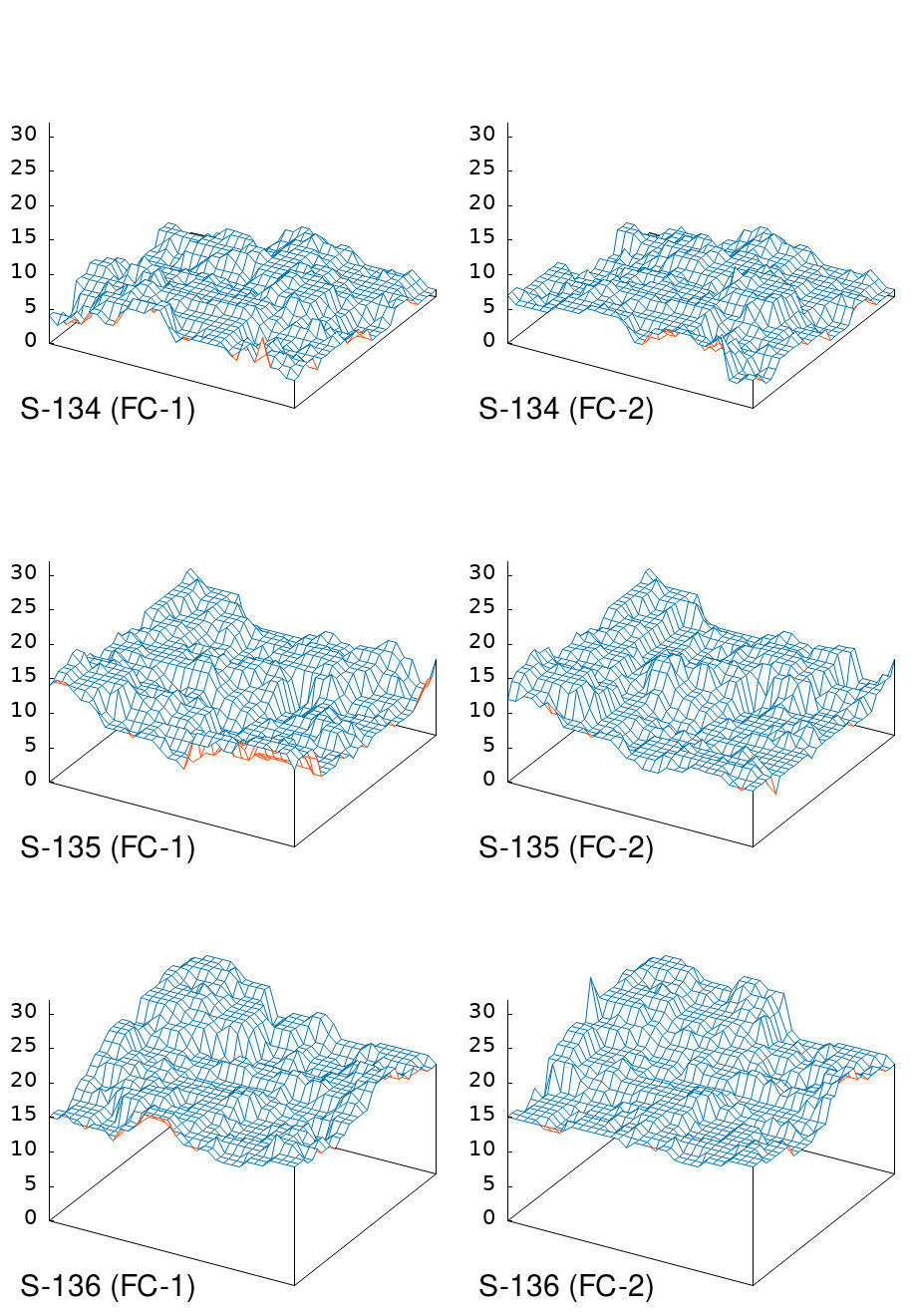}
\caption{Comparison of fracture surfaces obtained with different
	 fracture criteria. 
	 Three different samples are shown, {\rm S-134, S-135 
	 and S-136}.
	 {\rm FC-1} denotes the 'maximum normal stress' criterion, 
	 Eq.~(\ref{fcrit01}), and {\rm FC-2} denotes the 'maximum
	 shear stress' criterion, Eq.~(\ref{fcrit02}).
	 Fracture interfaces are red on the underside and
	 blue on top.
         \label{3x2cube_134b}}
\end{figure}

Previous calculations made with 
the 'original' criterion, Eq.~(\ref{f2m}), indicates a roughness 
exponent~$\zeta$ which varies considerably with the magnitude of the 
disorder~\cite{skj2}, i.e.,
$0.59<\zeta<0.78$. Here the smaller values~$\zeta\sim0.6$ 
correspond to strong disorder, $|D|\ge2$, and larger 
values~$\zeta\sim0.8$ to intermediate disorder, $D=1$.
These exponents are therefore somewhat high compared with the 
large-scale experimental result, $\zeta\simeq0.5$~\cite{boff,pons}. 
In Fig.~\ref{d1p5_fc0} we show the roughness
exponent obtained with the 'original' criterion~Eq.~(\ref{f2m}),
using $D=1.5$. Not surprisingly the result, $\zeta=0.72$, lies between the
results obtained for $D=1$ and $D=2$ in Ref.~\cite{skj2}, that is,
it lies between between $\zeta=0.78$ and $\zeta=0.62$. 
Using Eq.~(\ref{fcrit02}),
however, the value of the exponent reduces to the much lower value 
of $\zeta=0.52$, very close to the experimentally reported 
value for large length scales. 
This is notable in light of the fact that 
\begin{figure} [t]
\centering
\includegraphics[angle=0,scale=0.517]{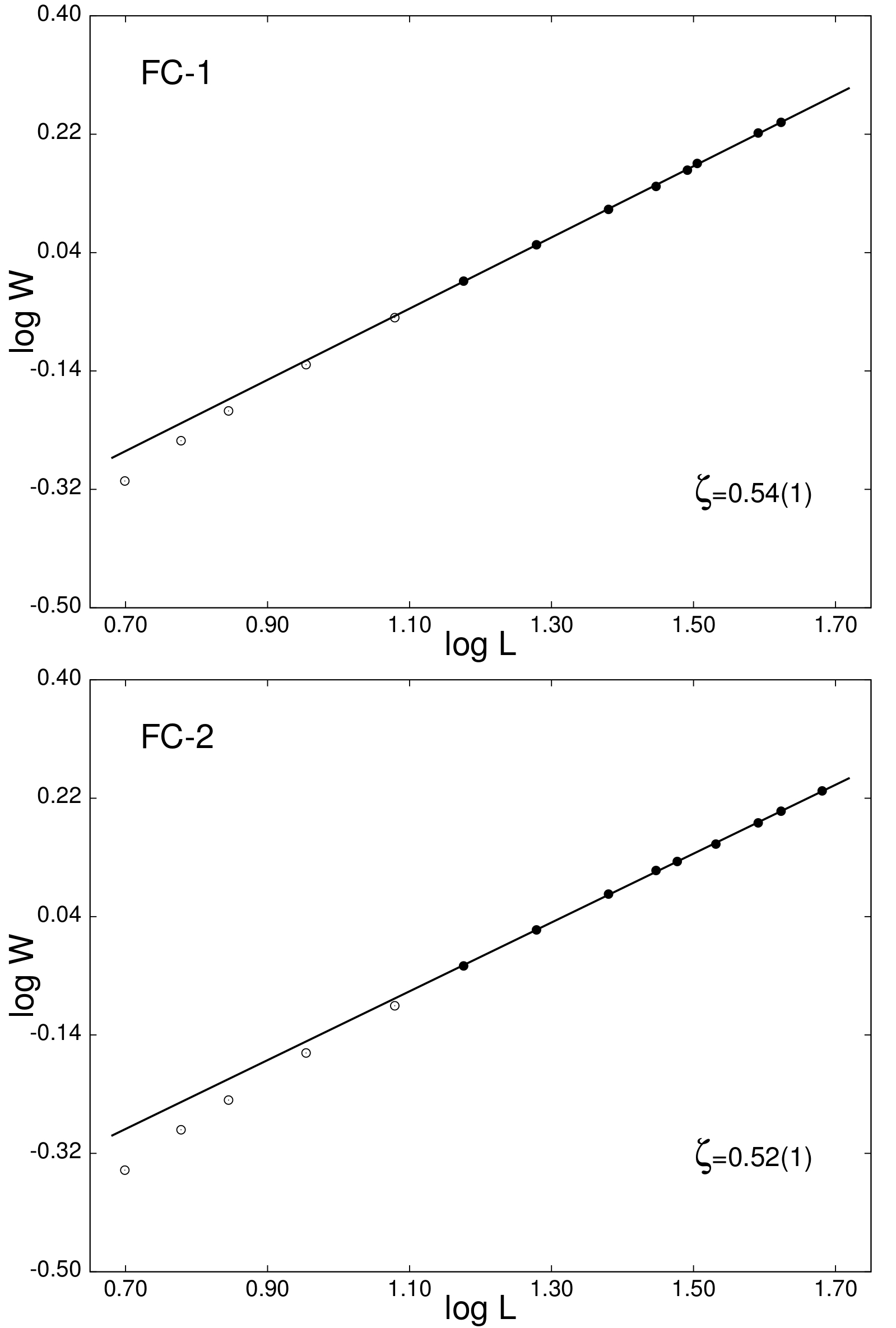}
\caption{Log-log plots for average roughness, $W$, as a function
         of system size, $L$, for a large number of
	 fractured samples. Disorder magnitude is $D=1.5$. 
	 Shown at the top {\rm (FC-1)} is the result obtained with 
	 the 'maximum normal stress' criterion, Eq.~(\ref{fcrit01}).
	 Shown below {\rm (FC-2)} is the result obtained with the
	 'maximum shear stress' criterion, Eq.~(\ref{fcrit02}).
         \label{d1p5}}
\end{figure}
the original criterion is wrong insofar as it only applies to
fracture with plastic deformations. 
Furthermore, the result obtained with Eq.~(\ref{fcrit01}),
$\zeta=0.54$, is similar to that obtained with Eq.~(\ref{fcrit02}).
Both results are shown in Fig.~\ref{d1p5}.
A comparison of fracture surfaces obtained with 
Eqs.~(\ref{fcrit02}) and~(\ref{fcrit01}) is shown in Fig.~\ref{3x2cube_134b}. 
The surfaces are seen to be very similar in this case,
close examination reveals only minor differences between the 
three samples.
Evidently, the inclusion of shear forces in the fracture
criterion fundamentally changes the detailed morphology of the crack 
surface. 

Brittle fracture on very small 
scales corresponds to the breaking of atomic bonds, thereby 
separating crystal planes. 
The resulting
fracture surface is flat until the crack front encounters 
an obstacle,
such as a grain boundary or a lattice defect. On small 
scales, therefore, it 
is not unreasonable to expect fracture surfaces akin to
those shown on the left in Fig.~\ref{3x2cube_134}. 
These were obtained with the 'erroneous' criterion, Eq.~(\ref{f2m}),
which, while lending much weight to axial force, has only 
weak contributions from bending and none from shear. 

Eq.~(\ref{f2m}) should, of course, not be regarded as an adequate 
criterion for brittle fracture on small scales. 
While the large scale fracture criteria, Eqs.~(\ref{fcrit02})
and~(\ref{fcrit01}), were derived from the theory of elasticity,
a small scale criterion would require
an analysis which takes into account the microscopic nature of the
structure, such as, for instance, binding by interatomic potentials.
From this a relevant functional form could be devised for
a fracture criterion to be used in discrete element modeling
at small scales. This should lend
appropriate weight to the tensile breaking which
gives rise to the cleavage process that takes place between crystal 
planes. At the same time it should include a less
dominating mechanism (based on bending or shear or both) 
that emulates encounters with grain boundaries and other discontinuities
within the crystal structure of the material.

\begin{figure} [b]
\centering
\includegraphics[angle=0,scale=0.517]{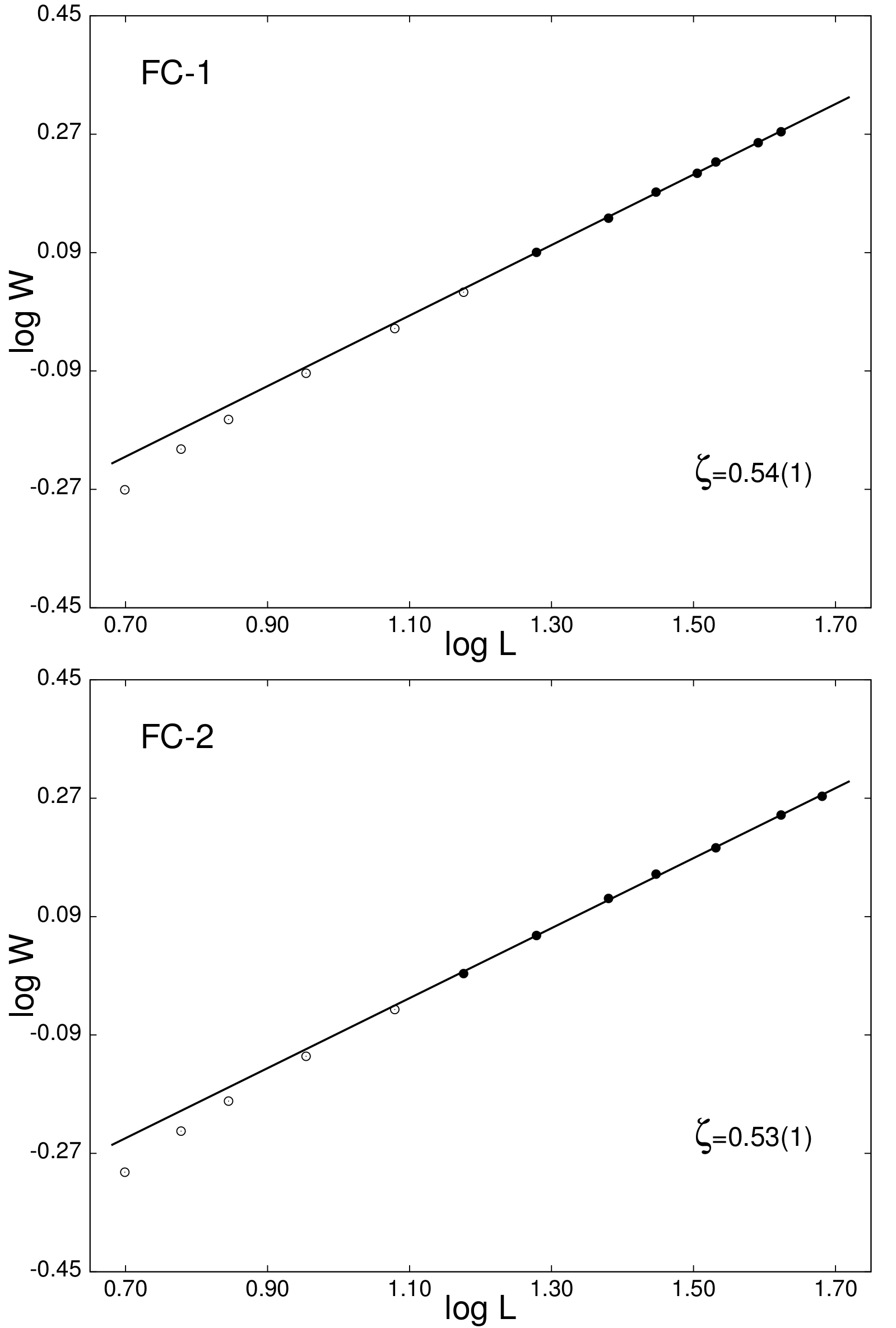}
\caption{Log-log plots for average roughness, $W$, as a function
         of system size, $L$, for a large number of
	 fractured samples. Disorder magnitude is $D=2$. 
	 Shown at the top {\rm (FC-1)} is the result obtained with 
	 the 'maximum normal stress' criterion, Eq.~(\ref{fcrit01}).
	 Shown below {\rm (FC-2)} is the result obtained with the
	 'maximum shear stress' criterion, Eq.~(\ref{fcrit02}).
         \label{d2p0}}
\end{figure}
This picture goes a long way towards explaining why two 
different exponents are obtained in the experiments. In numerical 
modeling with an appropriate fracture criterion which includes 
breaking due to shear, we obtain
$\zeta\sim0.5$ on scales large enough for shear to play an important role.
Contarary to this, $\zeta\sim0.8$ is expected on scales 
sufficiently small to be 
dominated by the crystal structure, as indicated by an 'erroneous'
criterion, such as Eq.~(\ref{f2m}), which lends a disproportionally
strong weight to tensile breaking.

In previous calculations with Eq.~(\ref{f2m}) it was seen that
the large scale roughness exponent $\zeta\sim0.5$ is approached
from above when the disorder strength is considerably increased, resulting in 
$\zeta=0.62$ being obtained for $D=2$ and
$\zeta=0.59$ for $D=4$~\cite{skj2}. The latter case, however, represents a
material structure with quite extreme variations in local strength
properties, perhaps unrealistically so for most materials. 

\begin{figure} [b]
\centering
\includegraphics[angle=0,scale=0.517]{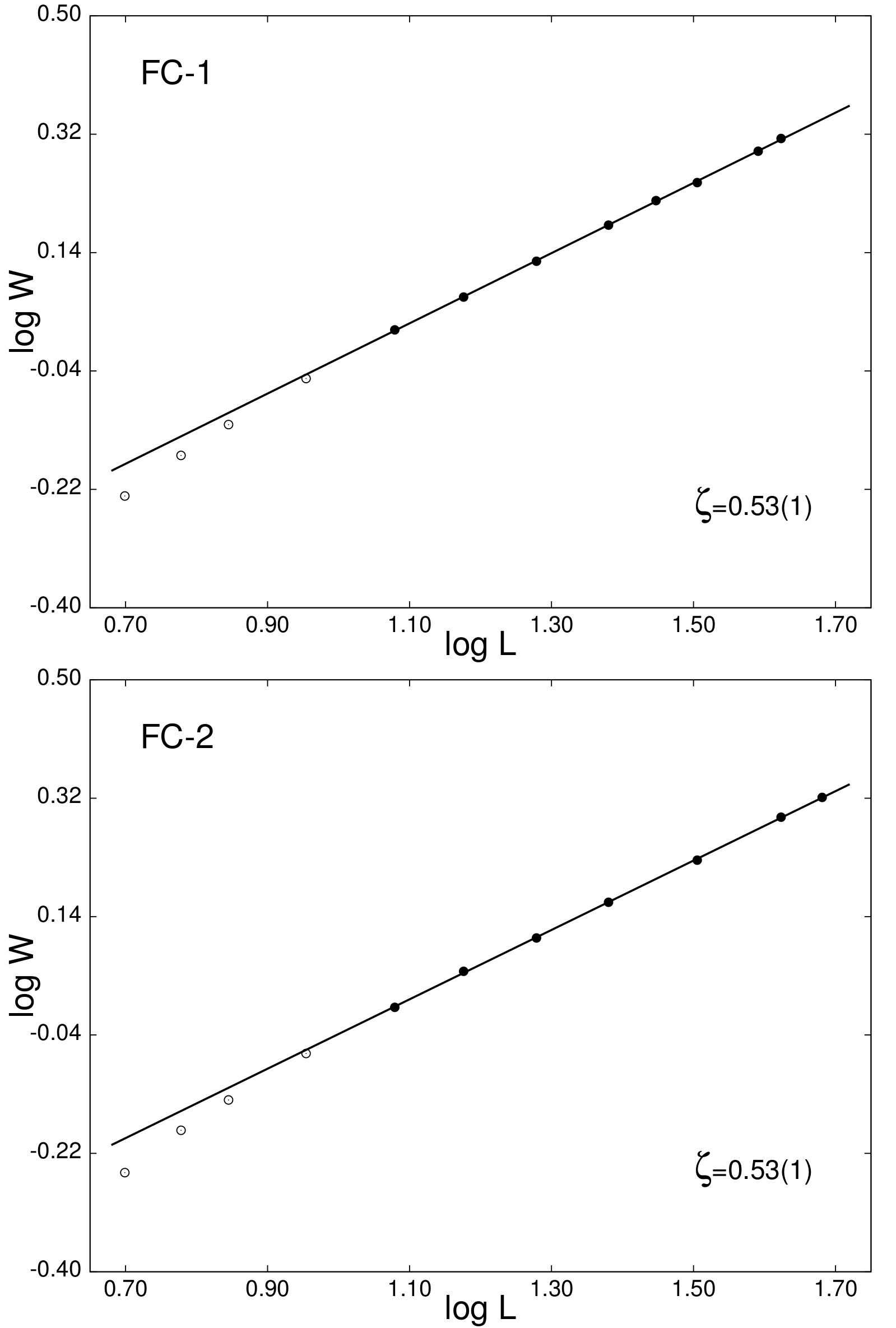}
\caption{Log-log plots for average roughness, $W$, as a function
         of system size, $L$, for a large number of
	 fractured samples. Disorder magnitude is $D=3$. 
	 Shown at the top {\rm (FC-1)} is the result obtained with 
	 the 'maximum normal stress' criterion, Eq.~(\ref{fcrit01}).
	 Shown below {\rm (FC-2)} is the result obtained with the
	 'maximum shear stress' criterion, Eq.~(\ref{fcrit02}).
         \label{d3p0}}
\end{figure}

\begin{figure} [t]
\centering
\includegraphics[angle=0,scale=0.50]{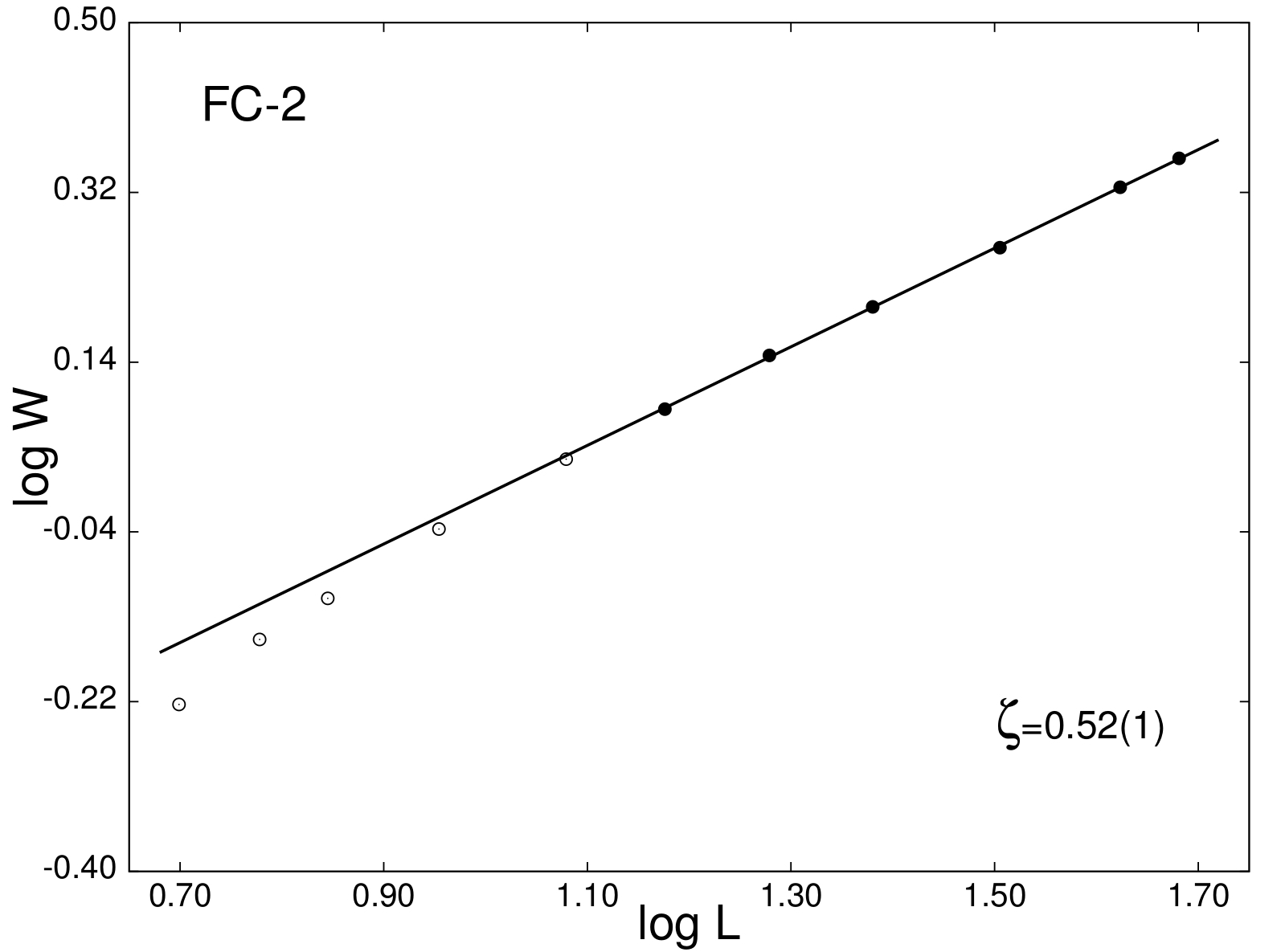}
\caption{Log-log plot showing average roughness, $W$, as a function
         of system size, $L$, for a large number of
	 fractured samples of each size. 
	 Disorder magnitude is $D=4$ and 
	 {\rm FC-2} denotes the result obtained with 
	 the 'original' fracture criterion, Eq.~(\ref{f2m}).
         \label{d4p0}}
\end{figure}

It is worth noting that with the new criteria, given
by Eqs.~(\ref{fcrit02}) and~(\ref{fcrit01}), the roughness
exponent remains in the vicinity of $\zeta=0.5$ for all 
disorders included in the present study. In other words, the
roughness appears to be universal with respect to disorder strength,
in contrast with what was found in Ref.~\cite{skj2}.
With $D=2$ the exponents obtained 
with Eqs.~(\ref{fcrit02}) and~(\ref{fcrit01}) 
are $\zeta=0.54$ and $\zeta=0.53$, respectively.
The result is shown in Fig.~\ref{d2p0}. At $D=3$ we obtain
$\zeta=0.53$ with both criteria, this is shown in Fig.~\ref{d3p0}.
Finally, at $D=4$ we obtain $\zeta=0.52$ using Eq.~(\ref{fcrit01}), in this
case we did not take the 
trouble to run an extra set of simulations for
Eq.~(\ref{fcrit02}). The result is shown in Fig.~\ref{d4p0}.
The apparently constant value which, within the uncertainties
of the straight-line fit, seems to fit all disorder 
strengths currently investigated with new fracture
criteria is $\zeta=0.53$. 

\section{Externally Applied Torque on a Cylindrical Shaft}
\label{cylshaft}
A typical property of brittle materials is that they are 
stronger in shear than in tension. As such, the criterion given 
by Eq.~(\ref{f2m}) does capture one essential feature of brittle 
fracture, namely the preference towards failure in axially 
tensile loading. If this was the only requirement a fracture 
criterion even more simple than Eq.~(\ref{f2m}) might have been 
sufficient, e.g., one that contains a single term only -- the 
ratio of the axial load to the failure load.

In discrete element modeling there is, however,
another feature which 
influences crack propagation and, ultimately, crack
morphology: the geometry of the lattice discretization.
Our current model is a cube lattice with nodes arranged as
shown in Fig.~\ref{6beam}. For a crack to propagate obliquely with
respect to the alignment of 'beams',
breaks will have to occur by lateral (transverse) deformation as 
well as by longitudinal (axial) deformation. For a cube lattice 
strained in the $Z$-direction lateral breaks are those that occur
within the $XY$-plane due to deformations transverse to the 'beam'
axis, i.e., shear deformations normal to (or bending deformations 
out of) this plane. In other words, a fracture plane which intersects the
$XZ$-plane at an angle of exactly 45 degrees will require an equal
number of transverse and longitudinal breaks. In localized fracture 
(very weak or no disorder)
these two types of breaking events should alternate as the line of 
intersection between the crack front and the~$XZ$-plane advances. 
\begin{figure*} [t]
\centering
\includegraphics[angle=0,scale=0.20]{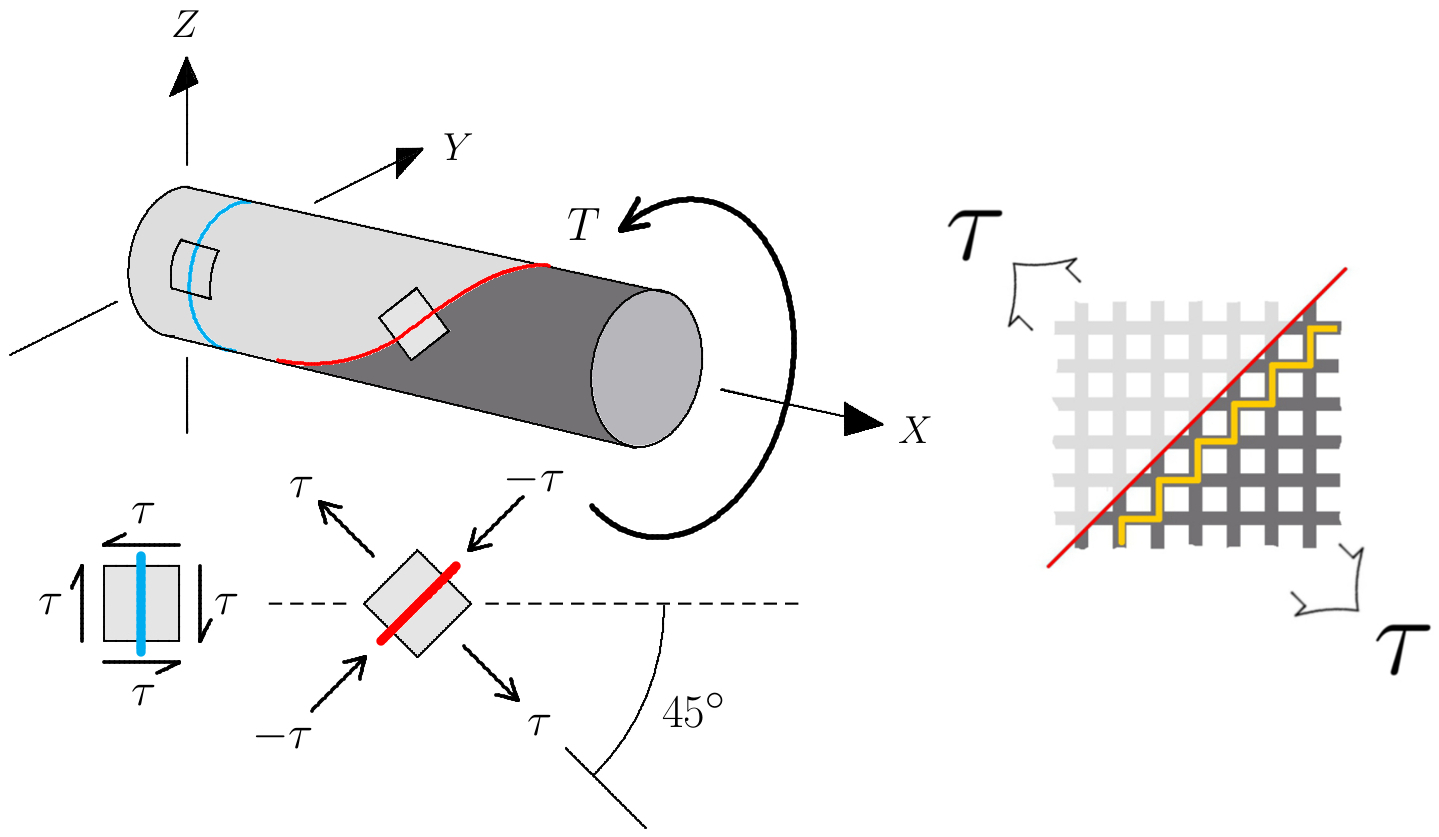}
\caption{A positive valued torque~$T$ is applied on the right
	 end of a cylindrical shaft, with the left end being held
	 fixed. Shown in blue is an intersecting
	 plane on which the shear stresses are at their highest. 
	 The associated material element is one of pure shear,
	 with the largest values obtained vertically or
	 horizontally.
	 Shown in red is an intersecting
	 helical surface perpendicular to which tensile forces
	 are highest. The associated material element
	 indicates the direction and value of the maximum tensile and
	 compressive normal stresses. This element is
	 contained within the lattice discretization on the
	 right, with a numerical realization of the helical surface 
	 depicted in yellow.
         \label{twist2}}
\end{figure*}
For a fracture plane which advances at a steeper angle, the ratio of 
horizontal to vertical breaks increases, while a more shallow fracture 
plane likewise requires relatively fewer horizontal breaks. 

Without 
providing for the possibility of shear failure, crack propagation 
would instead display a preference towards either the vertical or the 
horizonal plane,
depending on the direction of the external loading.
A situation requiring propagation along a plane inclined
at 45 degrees is the fracture of a cylindrical shaft due to
torque, see Fig.~\ref{twist2}. Directional inhomogeneity in
the elastic properties of the cylinder (other than disorder) could 
modify the angle of the fracture surface, but in the case of a shaft
with homogeneous material properties the emerging fracture angle
should be 45 degrees in brittle fracture. 
Any deviation from this should instead be
obtained by controlling the strength ratio of shear to tension in the
thresholds. For a discrete element model such a freedom of choice
is essential in order to obtain a crack which correctly reflects 
both the underlying structural disorder as well as other assumed material
properties.

\begin{figure} [t]
\centering
\includegraphics[angle=0,scale=0.37]{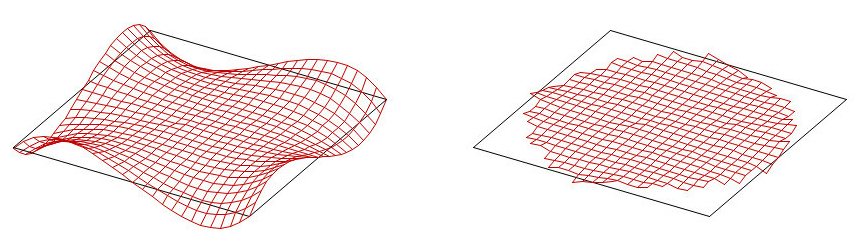}
\caption{Mid-point sections in a body subject to torsion. On 
	 the left is a square cross-section, showing the 
	 characteristic out-of-plane warping obtained for 
	 non-circular cross-sections. On the right is a circular 
	 cross-section
	 obtained by cutting beams outside a
	 circle inscribed within the square. 
	 Vertical displacements have been magnified by a factor of 50.
         \label{redsections}}
\end{figure}
We therefore next compare the new fracture criteria with the old criterion
by considering torsional fracture in a cylindrical shaft.
To construct such an object we first regard a rectangular column 
of discrete element 'beams' using the cube lattice discretization. 
From this a cylindrical body is obtained by inscribing a circle 
within the limits of the square cross-section, before cutting
away all 
discrete elements connected to nodes lying outside this circle.
This is shown in Fig.~\ref{redsections}, for a structure subject to
external torque. On the left the characteristic out-of-plane warping 
of non-circular cross sections is shown for a square cross-section
in the $XY$-plane,
\begin{figure} [b]
\centering
\includegraphics[angle=0,scale=0.24]{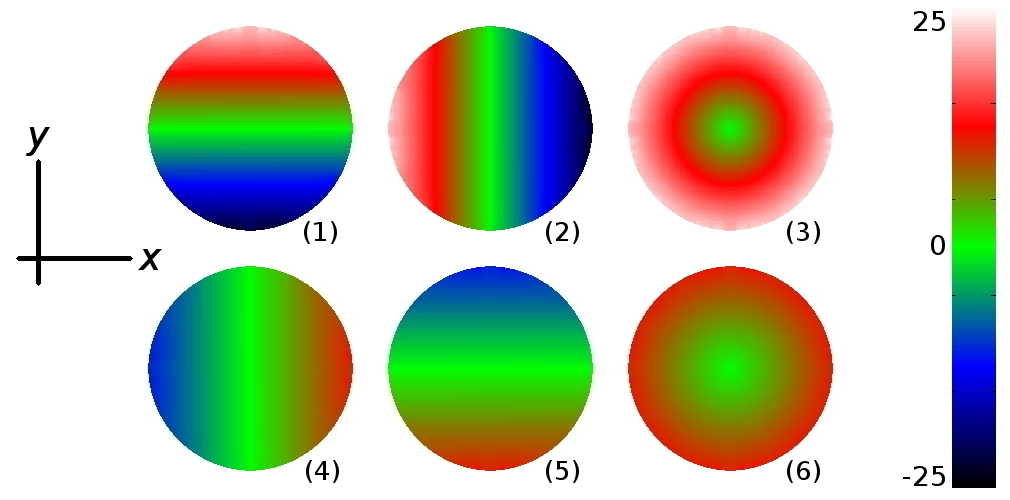}
\caption{Stresses on the central cross-section of a cylinder
         subjected to an external torque. Shown are
	 stresses in discrete element 'beams' which
	 connect horizontal layers in the cylinder, i.e.,
	 'beam'~$6$ in Fig.~\ref{6beam}.
	 Shear stresses are displayed at the top and
	 bending stresses at the bottom. See main text for
	 more information.
         \label{circstress}}
\end{figure}
while on the right a circular cross-section is shown. The amount
of torsion involved is the same in both cases, and the vertical
displacements have been exaggerated by a factor of fifty. 
Although it is unlikely that the warping of the square
cross-section influences the nature of the
fracture surface, we only consider circular cross-sections in the
following. (The very
minimal slanting seen at some of the edges of the circular cross-section
on the left in Fig.~\ref{redsections}
are probably finite-size lattice effects.)


\begin{figure} [t]
\centering
\includegraphics[angle=0,scale=0.34]{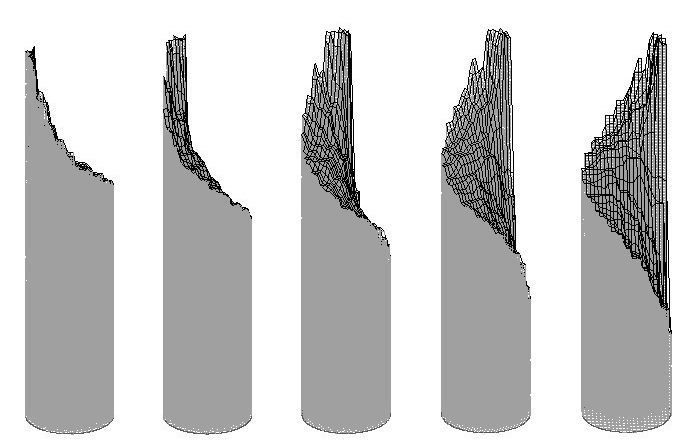}
\caption{A cylindrical shaft with diameter $d=25$ and height $H=201$
         which has been broken on the application of torque. The
	 fracture criterion used is Eq.~(\ref{f2m}). The
	 shaft is shown from five slightly different angles.
         \label{8xCyl_136}}
\end{figure}
\begin{figure} [b]
\centering
\includegraphics[angle=0,scale=0.31]{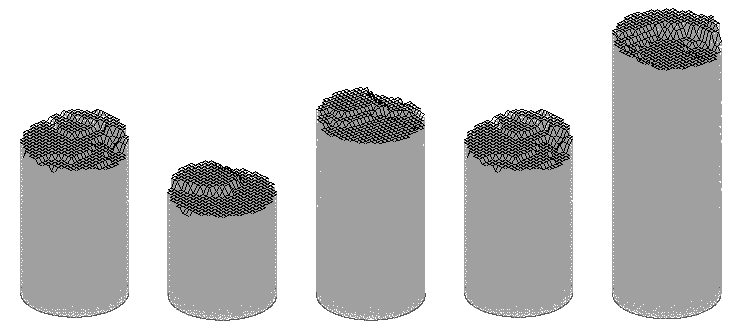}
\caption{Cylindrical shafts with diameter $d=25$ and height $H=101$
         broken by the application of torque. The
	 fracture criterion used is Eq.~(\ref{fcrit01}), where
	 tensile thresholds are twice as large as shear
	 thresholds. 
	 Five different samples have been shown here.
         \label{4xCyl_half5}}
\end{figure}
Shear and bending stresses on the cross-section 
of a cylindrical shaft induced by
torque is shown in Fig.~\ref{circstress}. At the top,~(1) and~(2) displays 
$X_{6}$ and~$Y_{6}$, i.e., shear forces in the $X$- and $Y$-directions.
These are obtained from the $j=6$ components of 
Eqs.~(\ref{forceXi}) and~(\ref{forceYi}), respectively.
Also shown, (3) is
\begin{equation}
    V_{xy}=\sqrt{X_{6}^{2}+Y_{6}^{2}},
    \label{sqrtVxy}
\end{equation} 
i.e., the bi-planar shear of Eq.~(\ref{VVVxy}). Below the shear
stresses are shown bending stresses. These, (4) and (5), are $V_{6}$
and~$U_{6}$, respectively. Also,~(6) represents
\begin{equation}
    M_{xy}=\sqrt{V_{6}^{2}+U_{6}^{2}},
    \label{sqrtMxy}
\end{equation}
or the bi-axial bending moment of Eq.~(\ref{maxcM}). This combines
bending within the $XZ$- and $YZ$-planes.
Fig.~\ref{circstress} shows that (with Eqs.~(\ref{matcons}
and~(\ref{epscons}) for the elastic constants)
shear stresses are about
twice as large as bending stresses. The necessity of using 
Eqs.~(\ref{sqrtVxy}) and~(\ref{sqrtMxy})
for consistency with rotational symmetry 
is also apparent.

Using the old criterion, Eq.~(\ref{f2m}), a typical example of 
the helical fracture surfaces obtained is shown from five slightly
different angles of rotation in Fig.~\ref{8xCyl_136}. The sample
has been subjected to a counter-clockwise rotation at the top
and a clockwise rotation at the bottom. 
All samples considered
currently have weak disorder, i.e., using $D=0.4$ in Eq.~(\ref{posd}). 
What is immediately apparent 
is that the angle of the fracture surface is rather steep. This is
a reflection on the fact that Eq.~(\ref{f2m}) is dominated by a
purely axial term while having only a weak contribution from bending.
It is this bending which provides the first local fractures since
the main forces are due to displacements transverse to the vertical axis. 
At some point, however, breaking due to horizontal tension becomes 
important. Crack propagation in the form of separation along vertical
lattice planes now becomes more dominant than breaking induced
by bending within the horizontal plane. The resulting fracture surfaces
tend to be very steep, significantly exceeding the 45 degree angle
in Fig.~\ref{twist2}, as is evident in Fig.~\ref{8xCyl_136}.

\begin{figure} [t]
\centering
\includegraphics[angle=0,scale=0.31]{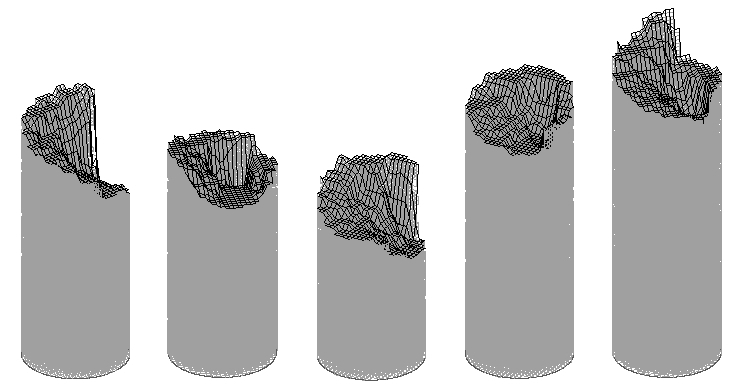}
\caption{Cylindrical shafts with diameter $d=25$ and height $H=101$
         broken by the application of torque. The
	 fracture criterion used is Eq.~(\ref{fcrit01}), where
	 tensile and shear thresholds are equal. 
	 Five different samples are shown.
         \label{4xCyl_equal5}}
\end{figure}
\begin{figure} [b]
\centering
\includegraphics[angle=0,scale=0.31]{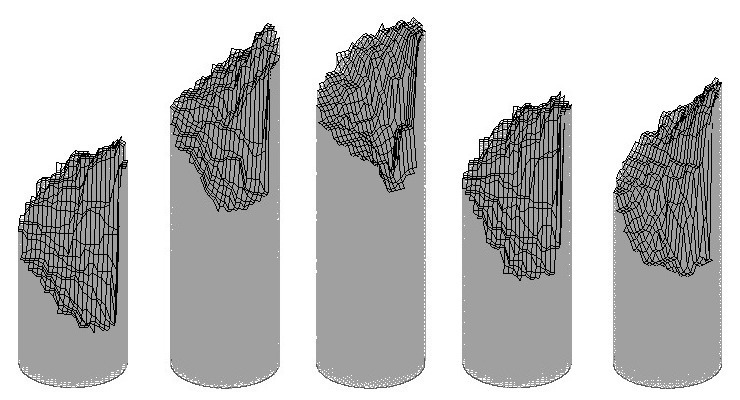}
\caption{Cylindrical shafts with diameter $d=25$ and height $H=101$
         broken by the application of torque. The
	 fracture criterion used is Eq.~(\ref{fcrit01}), where
	 shear thresholds are one and a half times as strong as 
	 tensile thresholds. 
	 The samples have been rotated to display more 
	 or less the same view.
         \label{4xCyl_1p5}}
\end{figure}
If instead we use our new criterion, Eq.~(\ref{fcrit01}), we have the
option to vary the strength relationship between shear and tension.
Assuming the material is stronger in tension than in shear we should
expect 'flat' fracture surfaces. Indeed, fracture surfaces obtained for a
shear/tension ratio of~$0.5$ are quite flat and five samples are shown 
in Fig.~\ref{4xCyl_half5}. Such a strength ratio is typical of many 
metals, including steel~\cite{deut}. These materials display less 
resistance towards the 
movement of dislocations within crystal 
planes and are thus more susceptible
to failure due to shear deformations.

\begin{figure} [t]
\centering
\includegraphics[angle=0,scale=0.31]{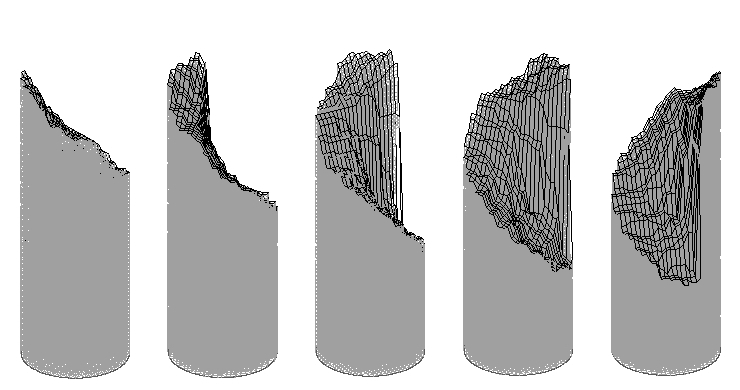}
\caption{A cylindrical shaft with diameter $d=25$ and height $H=101$
         broken by the application of torque. The
	 fracture criterion used is Eq.~(\ref{fcrit01}), where
	 shear thresholds are twice as strong as 
	 tensile thresholds. 
	 A typical sample is shown from five slightly different angles.
         \label{4xCyl_2p0}}
\end{figure}
\begin{figure} [b]
\centering
\includegraphics[angle=0,scale=0.31]{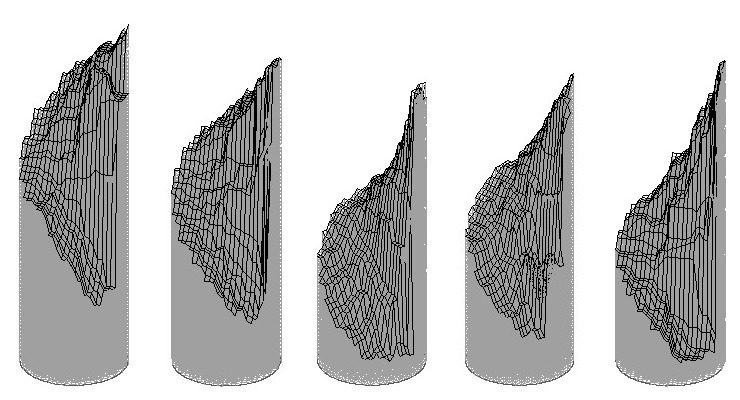}
\caption{Cylindrical shafts with diameter $d=25$ and height $H=101$
         broken by the application of torque. The
	 fracture criterion used is Eq.~(\ref{fcrit01}), where
	 shear thresholds are four times as strong as 
	 tensile thresholds. 
	 The samples have been rotated to
	 display more or less the same view.
         \label{4xCyl_4p0}}
\end{figure}
Increasing the stochastically generated shear strength to the 
point where it equals the stochastically generated tensile
strength changes the appearance of the fracture surfaces. Some of
the samples now display a slanting surface while others are reminiscent
of a cup-and-cone type surface (common in ductile fracture), see 
the five samples in Fig.~\ref{4xCyl_equal5}. 

Helical fracture surfaces of the type expected in Fig.~\ref{twist2}
appear as soon as the shear strength
is increased beyond the tensile strength.
In Fig.~\ref{4xCyl_1p5} shear
is one and a half times stronger than tension, with
five typical samples shown. 
A single sample where shear is twice as
strong as tension is shown from five 
slightly different angles
in Fig.~\ref{4xCyl_2p0}. Strength ratios where shear is stronger
than tension is typical of many rock types~\cite{jaeg}.

\begin{figure} [t]
\centering
\includegraphics[angle=0,scale=0.31]{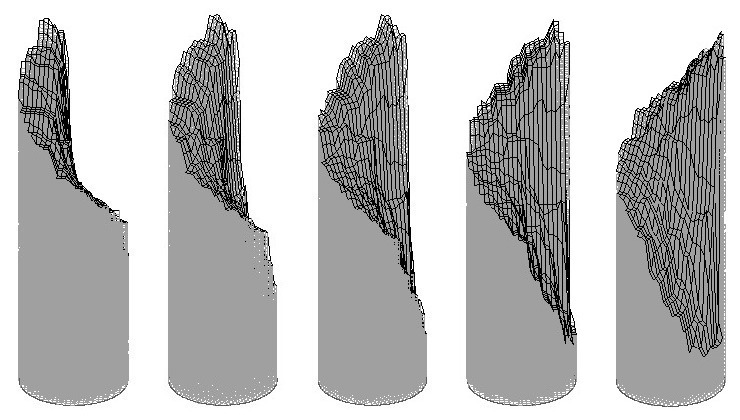}
\caption{Cylindrical shaft with diameter $d=25$ and height $H=101$
         broken by the application of torque. The
	 fracture criterion used is Eq.~(\ref{fcrit01}), where
	 shear thresholds are eight times as strong as 
	 tensile thresholds. 
	 A typical sample shown from five
	 slightly different angles.
         \label{4xCyl_8p0}}
\end{figure}

Further increase of shear strength relative to tensile strength
causes the angle of fracture to become progressively steeper.
In Figs.~\ref{4xCyl_4p0} and~\ref{4xCyl_8p0} the ratios are 
four and eight, respectively.

Even when compared with these rather extreme cases, however, fracture 
surfaces obtained with the original criterion,
Eq.~(\ref{f2m}), are even more steep. In
fact, they almost traverse the entire length of the sample. Such 
separation along 
vertical planes would perhaps be similar to the sort of fracture 
taking place when a broom stick is twisted until it breaks. Fracture 
then occurs as a separation of wood fibres that are parallel to the 
length axis of the shaft, 
rather than the breaking of such fibres 
in the direction normal to the axis, as would be expected in shear 
fracture.

\section{Concluding Remarks}
The choice of fracture criterion is shown to have a profound effect
on the crack morphology which obtains in calculations with a 
discrete element model for brittle fracture. The new fracture 
criteria are based on well known and long established relations 
and principles from the theory of elasticity, and replace a 
criterion which is really only relevant to plastic (rather than 
brittle) fracture. Modes of deformation such as axial strain, 
bending, shear and torsion are all included in the criteria used.

It is especially the inclusion of shear which most affects the
results obtained. Visually, the most conspicuous change is observed
at the weak end of the currently included range of disorders.
Here the resulting fracture surfaces appear considerably more 
rough. The way this influences the self-affine properties of the 
crack is to lower the roughness exponent to a value 
consistent with experimental findings, i.e., $\zeta=0.52$. 
What is more, the roughness exponent remains at this value for all 
disorders currently included, indicating a universal value.
An additional gain produced by allowing breaking in other
deformation modes, notably shear, is to enable the crack front to 
move more freely with respect to the lattice topology. Otherwise,
for a criterion with an axially dominant breaking mechanism, crack 
propagation will display a tendency to align itself in parallel 
with symmetry planes in the lattice. 
We have used a cube lattice, although this does not strictly
reproduce the correct macroscopic response to an external loading.
It is, however, less demanding on numerical resources than would 
be, say, an hexagonal close-packed lattice configuration.

No investigation into how results depend on the chosen elastic 
constants has been made in this study, this should probably
be addressed in a future study. 

\vspace{-5mm}
\begin{acknowledgments}
\vspace{-2mm}
This work was partly supported by the Research Council of Norway
through its CLIMIT and FRINATEK programmes, project numbers
199970 and 213462, respectively.
\end{acknowledgments}

\end{document}